\documentstyle[twocolumn,epsf,aps,prb]{revtex}
\begin{document}
\title{Dispersion of Ordered Stripe Phases in the Cuprates}

\author{R.S. Markiewicz}

\address{Physics Department and Barnett Institute, 
Northeastern U.,
Boston MA 02115}
\maketitle
\begin{abstract}
A phase separation model is presented for the stripe phase of the cuprates, 
which allows the doping dependence of the photoemission spectra to be 
calculated.  The idealized limit of a well-ordered array of magnetic and charged
stripes is analyzed, including effects of long-range Coulomb repulsion.
Remarkably, down to the limit of two-cell wide stripes, the dispersion can be
interpreted as essentially a superposition of the two end-phase dispersions,
with superposed minigaps associated with the lattice periodicity.
The largest minigap falls near the Fermi level; it can be enhanced by proximity
to a (bulk) Van Hove singularity.  The calculated spectra are dominated by
two features -- this charge stripe minigap plus the magnetic stripe Hubbard gap.
There is a strong correlation between these two features and the experimental 
photoemission results of a two-peak dispersion in La$_{2-x}$Sr$_x$CuO$_4$, and 
the peak-dip-hump spectra in Bi$_2$Sr$_2$CaCu$_2$O$_{8+\delta}$.   The 
differences are suggestive of the role of increasing stripe fluctuations.  
The 1/8 anomaly is associated with a quantum critical point,
here expressed as a percolation-like crossover.  A model is proposed for the 
limiting minority magnetic phase as an isolated two-leg ladder.
\end{abstract}


\narrowtext

\section{Introduction}

Evidence for stripe phases in the cuprates continues to grow.
Particularly in the La$_{2-x}$Sr$_x$CuO$_4$ (LSCO) family, a convincing case
for (predominently dynamic or disordered) stripes can be made, based on 
elastic and inelastic neutron scattering\cite{Tran,Yam,Suz,Birg,BirgLSO}, 
NMR and NQR\cite{Imai,Jul}.
In other systems, the evidence is more ambiguous.  In YBa$_2$Cu$_3$O$_{7-\delta}
$ (YBCO), there is now\cite{M1,Arai} clear evidence for incommensurate 
modulation of the inelastic magnetic neutron scattering near $\vec Q=(\pi ,
\pi)$, but so far only in underdoped YBa$_2$Cu$_3$O$_{6.6}$.
Balatsky and Bourges\cite{B1} find a broad commensurate
peak, but the width of the peak scales with doping in exactly the same way as
the incommensurability in LSCO, suggestive of an unresolved underlying
incommensurability in YBCO, as well.  Also, de Lozanne\cite{EdL} finds direct
STM evidence for incommensurate modulations (parallel to the chains) with a
similar periodicity to the neutron data.  Mook\cite{M2} has reported similar 
incommensurate neutron peaks in Bi$_2$Sr$_2$CaCu$_2$O$_{8+\delta}$ (BSCCO).
Potentially stripe-related phonon anomalies have been reported in both 
LSCO\cite{Ega} and YBCO\cite{M3}.  Doping with Zn seems to stabilize the stripe
phase\cite{ANO}.  Photoemission evidence\cite{Bian,SWK} for stripes has been 
controversial\cite{Camp,ZX2}.

Over the same doping regime, there is also evidence for a pseudogap, and it is
an important problem to understand how both pseudogap and stripes can coexist.
In particular, photoemission finds a dispersion consistent with the 
two-dimensional (2d) energy bands, whereas in the stripe phase the magnetic 
stripes should be insulating, leading to a one-dimensional (1d) dispersion 
along the charged stripes.

The presence of stripe phases raises important issues of how energy dispersion 
and even Fermi surfaces can be well defined concepts in the presence of 
fluctuating stripes.  An important insight into this problem is the finding by 
Salkola, et al. (SEK)\cite{SEK} that a well-defined average dispersion persists 
even in the presence of strongly fluctuating stripe order.  
However, in that paper, and a related calculation\cite{Sei}, the stripes were
modelled by a charge density wave like order, with sinusoidally varying hole
density.  Several unrestricted Hartree-Fock\cite{HF,VLLGB,Yon,MVz} or slave
boson\cite{Sei1} calculations find evidence for much sharper density variations.
The present paper analyzes a phase separation scenario, modelling the stripes 
as associated with free energy minima at two characteristic hole densities. 
This allows the doping dependence of the stripes and the resulting 
photoemission spectrum to be analyzed.  

It is found that long-range stripe order can persist even in the presence of 
Coulomb interactions.  The resulting dispersion is clearly recognizable as a
superposition of the magnetic and charged stripe dispersions, with
superimposed minigaps due to the stripe order.  These dual dispersions provide a
natural interpretation for the experimentally observed photoemission 
dispersions, tying together results on LSCO, BSCCO, and Sr$_2$CuO$_2$Cl$_2$ 
(SCOC).  In the model, the 1/8 anomaly can be understood as a form of quantum
critical point (QCP), associated with a crossover between a magnetic stripe
dominated regime and a charged stripe dominated regime.  Within the latter
regime, the $(\pi ,\pi )$ spin gap in YBCO is related to the behavior of a
two-leg ladder (isolated magnetic stripe).

Remarkably, within the charge stripe dispersion, a clear signature of the
two-dimensional Van Hove singularity (VHS) persists, down to the limit of a
single, two-Cu wide stripe.  There is a strong coupling of the minigaps 
with this VHS, leading to a novel {\it stripe-induced VHS splitting}.  The
doping dependence of this splitting closely resembles that of the pseudogap.

The paper is organized as follows.  Section II shows that a low hole doping of
the charged stripes, $x_0\sim 0.25$ is not only compatible with experiment, but
also makes sense theoretically, in terms of kinetic-energy stabilized stripes.
The models for the magnetic and the charged stripes are introduced in Section 
III, along with a discussion of long-range Coulomb interaction.  Section IV
gives the results of the stripe calculations, which self-consistently determine 
the hole distribution.  The doping dependence of the dispersion is presented, 
for varying strengths of Coulomb 
repulsion.  Finally, the effect of an additional (ferromagnetic) interaction
on splitting the VHS degeneracy on the charged stripes is discussed.  In
Section V, these results are compared to experiment, and a consistent model of
the photoemission in LSCO and BSCCO is presented.  Section VI points out that 
the model has a QCP -- actually, a series of `magic doping' QCP's, of which
the simplest is the 1/8 anomaly.  To illustrate the resulting crossover, an
additional calculation is presented in Section VII, showing how the doping
dependence of the spin gap in YBCO can be understood.  
Possible explanations are also presented for the saturation of the 
incommensurability $\delta$ vs $x$ found by Yamada, et al.\cite{Yam}

Section VIII includes discussions of the interpretation of the peak-dip-hump
structure in BSCCO, the new stripe-VHS spitting pseudogap, a discussion of
Fermi surfaces and remnant Fermi surfaces in the stripe phase, and comparison
with earlier calculations.  A summary of the principal conclusions of this
work is given in Section IX.

\section{Fractionally-Occupied Stripes}

\subsection{Comparisons with other Oxides}

Stripe arrays have now been found in a number of oxides, most notably nickelates
and manganites.  The similarities of cuprates with nickelates are particularly 
close: in
both systems, the charged stripes act as antiphase boundaries for the magnetic
stripes, and in both, the charge order arises at higher temperature than the
magnetic order\cite{Tran,Cfirst}.  The nickelate stripes run diagonally 
(with respect to the Ni-O-Ni bonds); this is also true of the LSCO stripes, in 
the spin glass regime\cite{diag}, $x\sim 0.04-0.06$.  However, in the
superconducting regime, $x>0.06$, the cuprate stripes are generally 
horizontal and vertical.
\par
One striking difference is that in the nickelates and manganites, the charged
stripes correspond to integer doping (one hole per Mn or Ni), leading to simple
patterns\cite{nipat,mnpat} of commensurate stripe arrays. There are prominent 
phase transitions at rational fractions, $1/2$, and $1/3$, corresponding to 
holes on every $n$th row, with evidence for commensurability locking in between 
(i.e., the 1/3 phase persists in an extended doping range about $x=1/3$.)  
Consistent with integer filling, the phases are all insulating\cite{nick}.  In 
contrast, in the cuprates the phases are all conducting or weakly localized, and
the only fraction which appears prominently is 1/8.  

In the present paper, a simple explanation is proposed for this distinction.
The charged stripes are fractionally doped, with approximately 1/4
hole per Cu (hence explaining the finite conductivity).  The magical 1/8 doping
would then correspond to the simplest `commensurate' pattern of these stripes.  
\par
The stability of the stripe phase decreases in the sequence manganites, 
nickelates, cuprates.  Thus, while there are beautiful electron microscopic
images of long-range stripe order in the manganites\cite{mnpat}, stripes in
the cuprates are mainly fluctuating, with only short-range order.  Within the
present model, this pattern is readily understood, since the charged stripes
are stabilized by CDW instabilities; this is similar to models for the 
nickelates and manganites\cite{MK}.  The strength of this instability can be 
estimated by comparing the strength of electron phonon coupling, which follows 
the same sequence: manganites (with well-defined Jahn-Teller polarons), 
nickelates\cite{BE}, cuprates.  It is only in the cuprates where the interaction
is so weak that a fractional occupation can be stabilized, 
and it is only in the cuprates that the stripe formation is so weak that
superconductivity can successfully compete.

\subsection{Origin of Fractional Occupation}

Hartree-Fock calculations\cite{HF} of the tJ model find that the holes condense 
onto domain walls between antiferromagnetically ordered domains, producing fully
occupied charge stripes -- one hole per Cu.  
However, neutron diffraction\cite{Tran} finds a charge modulation of
periodicity four Cu atoms at $x=0.125$, which implies only 1/2 hole per cell.
Tranquada, et al.\cite{Tran} suggested a model for the charged stripes, based on
their experience with stripes in nickelates.  The hole-doped stripes are one 
cell wide, and have a hole on every other site.  A microscopic model for such a 
domain wall can be derived\cite{Zaa3} by incorporating a charge-density wave 
(CDW) instability along the stripes, treating them as one-dimensional metals.
However, such states with integral hole 
doping are likely to be insulators, as is the case in the stripe phases of the 
nickelates\cite{nick}, whereas the cuprates are either conducting or weakly 
localized.  

Moreover, fractional hole occupation would seem to be more natural for the tJ 
and Hubbard models, since the energy of doped holes is lowered by finite 
hopping $t$ in a partially filled band.  Visscher\cite{Vis} and 
Nagaev\cite{Nag} showed that the holes enhance their kinetic 
energy by creating local ferromagnetic domains (ferrons) in which they are free 
to hop.  This leads to a preferred hole density, $x_f$ inside the ferron domain.
In a two-dimensional, tJ version of the model (letting $\hbar^2/2m\rightarrow 
ta^2$, with $a$ the lattice constant),
\begin{equation}
x_f=\sqrt{zS^2J\over\pi t}\simeq 0.334,
\label{eq:1z}
\end{equation}
with $z=4$ the number of nearest neighbors of a given Cu, and I have assumed
$J/t=0.35$.  A similar result was found by Nayak and Wilczek\cite{NaW}.
Nagaev's model is a large-$S$ theory, and Emery and 
Kivelson\cite{EmK} extended it to $S=1/2$, although they did not address the 
issue of $x_f$.  Auerbach and Larson\cite{AuLa} showed that a single doped 
$S=1/2$ hole will spread out over a ferromagnetic domain covering 5 lattice 
sites, suggesting a comparable value for $x_f$, $~>0.2$ holes per cite, on 
average (since the hole has a higher probability of being on the central atom). 
Recent density matrix renormalization group (DMRG) calculations of the tJ model 
undertaken by White and Scalapino (WS)\cite{WhiSc,WhiSc2} find charged stripes
which are two Cu's wide, with an average hole doping of $~0.25$ hole per Cu on 
the charged stripes.  These calculations are further discussed in Appendix A.  

Since the charged domains are stabilized by the hole kinetic energy, it is
plausible that enhancing the kinetic energy could enhance the stability of the
hole-doped stripes.  Thus, in a generalized Hubbard model, with next-nearest
neighbor hopping $t^{\prime}$, it is found that a macroscopic ferromagnetic 
phase is stabilized in the vicinity of the Van Hove Singularity (VHS)\cite{FMV}.
Moreover, an extended Hartree-Fock analysis\cite{MVz} finds phase separated
states smoothly evolving between the AFM and FM regions, from a single magnetic
polaron to FM stripes to a uniform FM phase.

However, such ferromagnetic domains have not been observed in the cuprates.
Nevertheless, there are alternative VHS routes to fractionally-occupied stripes.
The large density of states (dos) associated with a VHS can drive a large number
of competing electronic instabilities\cite{Sch2,MarV}, and it was early 
suggested that this could be the origin of nanoscale phase separation in the 
cuprates\cite{RM3}.  In particular, it was demonstrated that strong
electron-phonon coupling could stabilize a charge-density wave phase near the
VHS\cite{RM3,Pstr}.

\subsection{Viability of VHS Models}

In any model of stripe phase formation based on Fermi surface features, there 
is a fundamental question of self-consistency: do the features persist in the
limit of an isolated stripe?  Can one still recognize bulk features of the band
structure and Fermi surfaces of the phases forming the stripe array?
This is one of the main issues that this paper resolves: even in the limit of
nanoscopic stripes, the band structure is recognizably a superposition of the
structures of the two end phases.  The main role of stripe order is to introduce
miniband gaps into this structure.  

In the particular case of the VHS's, there were a number of preliminary 
indications which suggested such an affirmative answer.  First, SEK\cite{SEK} 
found that an average 
dispersion persists in the presence of fluctuating stripes; the resulting `flat 
bands' are a signature of the VHS.  Secondly, within a group theoretical (SO(6))
model\cite{MarV}, the Van Hove instabilities all remain well-defined on a single
plaquette of 2$\times$2 Cu atoms, so {\it a fortiori} they should remain well 
defined on a 2-leg ladder.  Indeed, Lin, Balents,
and Fisher\cite{LBF} found an SO(8) group controlling the physics of the
2-leg ladder.  When one eliminates\cite{Rome} certain one-dimensional operators 
(which break the $k\rightarrow -k$ symmetry along the ladder), one is left with 
the same SO(6) group introduced earlier for the VHS.  Such a correspondence 
would fail for a single-leg ladder.  
\par
Hence, the present model is restricted to stripes which are an even number of
cells (or Cu atoms) wide.  This point was previously postulated for the 
magnetic stripes, in terms of spin gaps associated with even-legged 
ladders\cite{TTR}.  Moreover, WS find two-Cu wide charged stripes in their DMRG 
calculations\cite{WhiSc}.  With this assumption, it is found that a VHS-like
feature can be clearly resolved near the Fermi level in the stripe phases.
Moreover, the stripes provide a new mechanism for VHS splitting -- minigaps --
which can generate a pseudogap with the correct doping dependence.

\section{Modeling the Stripes}

While the stripes are likely to be strongly fluctuating, the band structure 
modifications should be strongest, and can be analyzed in most detail, in an 
ordered stripe phase.  Hence, the present calculation assumes perfectly ordered 
stripe phases to describe this `worst case' scenario.  
It will be assumed that there are two
preferred hole densities, $x\sim 0$ on the magnetic stripes, and $x_0\sim$ 
0.25 holes per Cu on the hole-doped stripes.  Coulomb effects lead to 
additional charge relaxation, and a more uniform distribution of charge, 
Section III.D.

\subsection{Model for the Magnetic Stripes}

In the insulating phase, a variant of the spin-density wave (SDW) model 
studied by Schrieffer and coworkers\cite{SWZ,KaSch} is used.  This model works 
surprisingly well in the large-U limit\cite{BSW}, reproduces the spin wave
spectrum of the Heisenberg model, and has served as the basis for a number of
extended treatments of correlation effects\cite{ChM,SPS,MiG,GLLV}.  For 
realistic parameters ($t$, $t^{\prime}$, $U$), the model has a Mott-Hubbard gap 
of 2eV, and can reproduce the dispersion found in the 
oxyclorides\cite{Well,oxy}, Appendix B.

\begin{figure}
\leavevmode
   \epsfxsize=0.33\textwidth\epsfbox{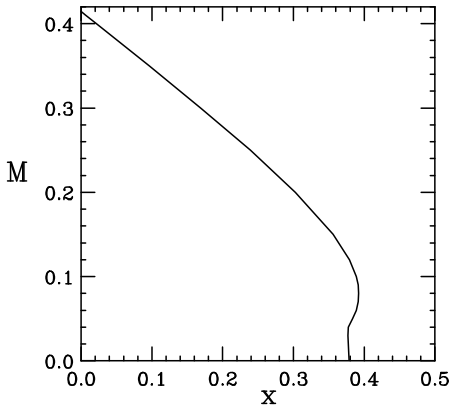}
\vskip0.5cm 
\caption{Doping dependence of magnetization in the SDW model.}
\label{fig:1}
\end{figure}

The dispersion of the one-band model can be written
\begin{equation}
\epsilon_k=-2t(c_x+c_y)-4t^{\prime}c_xc_y,
\label{eq:A0}
\end{equation}
with $c_i=\cos{k_ia}$.  Writing $\epsilon_{\pm}=(\epsilon_k\pm\epsilon_{k+Q})
/2$, the eigenvalues in the presence of a Hubbard U become
\begin{equation}
E^{\pm}=\epsilon_+\pm\sqrt{\epsilon_-^2+\bar U^2},
\label{eq:A1}
\end{equation}
where $\bar U=Um_Q$.
In the limit $\bar U>>t$, the lower Hubbard band may be approximated 
\begin{equation}
E^-=-\bar U-4t^{\prime}c_xc_y-J(c_x+c_y)^2,
\label{eq:A2}
\end{equation}
with $J=2t^2/\bar U$.  The parameters can be determined by fitting to the
observed photoemission dispersion in SCOC.  For simplicity, one can use
analytical expressions for the parameters at three $k$-space points: $E^-(\pi
/2,\pi /2)=-\bar U$, $E^-(\pi ,0)=-\bar U+4t^{\prime}$, $E^-(0,0)=-4t^{\prime}
-\sqrt{\bar U^2+16t^2}$ (Eq.~\ref{eq:A2} is not sufficiently accurate for this
purpose).  The fit yields $t=325 meV$, $\bar U/t=2.5$, and $\tau =2t^{\prime}/t=
-0.552$.
Solving the gap equation at half filling, this value of $\bar U$ corresponds to
$U/t=6.03$, $M_Q(x=0)=0.414$ (Fig.~\ref{fig:1}), or 83\% of the classical value.
\par
For these parameters, $M(x)$ is multivalued for $x\ge 0.38$.  This implies that
the magnetic to non-magnetic transition is first order.  This is discussed 
further in Appendix B.  However, this density is rather higher than expected 
for charged stripes.  In LSCO, the VHS splitting seems to terminate
near $x=0.26$\cite{Lor,Surv}, and similar results are found below for YBCO.  
In a number of models\cite{Pstr,FMV}, the AFM instability is replaced by a
second instability, driven by splitting the VHS degeneracy.  Note that the bare
($U=0$) VHS falls at $x=0.25$ for $\tau=-0.559$, very close to the value needed 
to explain the dispersion in the insulating phase.

If $M_Q$ is interpreted as the long-range antiferromagnetic order parameter,
then the model does a poor job in describing the temperature and doping 
dependence of the N\'eel transition\cite{DJK}, $T_N$, yielding $T_N\sim U/4$.  
Figure~\ref{fig:1} shows that, while $M_Q$ is strongly renormalized by doping,
the mean-field theory underestimates the rapidity of the falloff of $T_N$ with
x. However, the mean-field results are best reinterpreted as representing
{\it short-range order} -- the magnetic fluctuations -- and hence the 
renormalization of the {\it splitting} into upper and lower Hubbard bands.
In this case, the mean-field calculations are in good agreement with exact
diagonalization calculations\cite{EsO}.
The fact that the gap is much smaller in the doped phase is consistent with the 
experimental observation\cite{exp} that the upper Hubbard band rapidly
disappears with doping. 

\subsection{Model for the Charged Stripes}

It is assumed that the hole-doped stripes are stabilized by splitting the
VHS degeneracy, at the doping $x_0\sim 0.25$ where the VHS falls at the Fermi
level.  An earlier slave boson calculation\cite{Pstr} demonstrated that
electron-phonon coupling could provide that stabilization energy, even in the
presence of strong correlation effects.  A ferromagnetic interaction\cite{FMV}
can produce similar splitting.

While the earlier electron-phonon calculation involved a three-band model, 
here a simpler one-band model will be adopted.  A parametrized form of the
free energy vs doping found in the self-consistent calculation\cite{Pstr}, 
Fig. \ref{fig:16}, will be assumed, to stabilize the stripe phase.  It is
convenient at present to {\it not} introduce any mechanism to split the
VHS degeneracy.  This allows a definitive answer to an important question: can
evidence for the VHS still be found in the presence of a well-defined stripe
phase?  The answer is a clear yes: the resulting dispersion is a superposition 
of the magnetic dispersion and the charged stripe dispersion, with recognizable
VHS feature.  What is more, the stripe phase minigaps provide a {\it new
mechanism of VHS splitting}, with a doping dependence comparable to the
experimental pseudogap.

A very simple doping dependence of the parameters is assumed.  From 
Eqs.~\ref{eq:A0}-\ref{eq:A2}, for finite $U$
$t$ is renormalized by a factor $t/Um_Q$, Fig.~\ref{fig:1}, so the increase of 
$t$ with doping is accomplished by the decrease in $m_Q$, the ordered moment.  
We will thus make a simple {\it Ansatz} that the only effect of doping is to 
renormalize 
\begin{equation}
m_Q\rightarrow m_Q(1-x/x_0).
\label{eq:1}
\end{equation}
Since the stripes are predominantly near the limiting states $x=0$, $x_0$, the
detailed nature of the intermediate states is relatively unimportant.  
As noted above, Eq. \ref{eq:1} neglects the gap on the charged stripe; 
in Section IV.D, a ferromagnetic interaction will be included on the charged
stripes, to show that the VHS splitting is preserved in the striped phase.

\subsection{Free Energy Minima}

\begin{figure}
\leavevmode
   \epsfxsize=0.33\textwidth\epsfbox{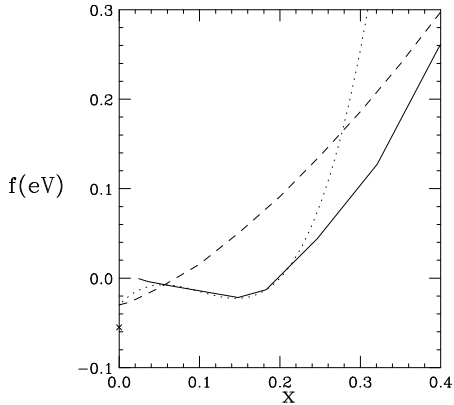}
\vskip0.5cm 
\caption{Free energy calculated in a three-band slave boson 
theory\protect\cite{Pstr}, showing phase separation between a flux phase (dashed
line) and a CDW phase, with $V_{ep}=1eV$ (solid line).  Dotted line is a fit to
Eq.~\protect\ref{eq:1c}.}
\label{fig:16}
\end{figure}

To stabilize the stripe densities at the values $x=0$ for magnetic stripes, and
$x=x_0=0.25$ for the charged stripes, the following free energy is introduced,
based on the results of slave boson calculations for the three-band 
model\cite{Pstr}:
\begin{equation}
f_0(x)=\mu_0x(1-{x\over x_0})^2,
\label{eq:1c}
\end{equation}
for $x>0$.  (At $x=0$ there is a cusp-like minimum, associated with the Mott 
gap in the chemical potential.)  Figure~\ref{fig:16} illustrates the free
energies calculated in a three-band slave boson calculation for competing
magnetic (here a flux phase -- dashed line) and charged (CDW -- solid line)
phases, and a fit of these to Eq.~\ref{eq:1c}, with $\mu_0=0.9eV$.  For these
calculations, $x_0=0.16$ was assumed.  This should be compared to Fig. 2 of 
Ref.~\onlinecite{Pstr}. Note that there is an error in the
caption of that figure: the CDW phase there corresponds to a weaker coupling, 
$V_{ep}=0.6eV$.  

Equation~\ref{eq:1c} is a convenient form for parametrizing the confining 
potential of a striped phase.  It has only two parameters, $x_0$ and $\mu_0$,
or equivalently, $f_m=4\mu_0x_0/27$, the maximum free energy barrier, at $x_0
/3$.  In the present stripe phase calculations, these parameters are taken as 
$x_0=0.25$ and $\mu_0=0.312eV$, or $f_m=11$meV. This value corresponds to $V_
{ep}=0.6eV$ of Ref.~\onlinecite{Pstr}, and allows us to see that even a 
relatively modest confining potential can stabilize the stripe phase against the
Coulomb potential.  This free energy corresponds to an additional chemical 
potential
\begin{equation}
\mu (x)=-\mu_0(1-{x\over x_0})(1-{3x\over x_0})
\label{eq:1d}
\end{equation}
for $x>0$.  In the calculations, this $\mu (x)$ is added to the potential on 
each row, and the local density adjusted until self-consistency is attained.

At $x=0$, $\mu$ has a discontinuity, the Mott-Hubbard gap.  Hence, at this 
point, the Fermi level can take on any value inside the gap.  To model this in
a computationally stable manner, the discontinuous step in $\mu$ is replaced by 
a linear ramp, connecting the values of $\mu$ at $x=-0.01$ and $x=+0.01$, and 
assuming $\mu (x=-0.01)=-\mu (x=+0.01)$.  Thus, when the calculation finds $|x|
<0.01$, it generally implies that the Fermi level is in the gap of the magnetic 
stripes.  However, due to hybridization with holes in the charged stripes, it
is possible to have a well-defined Mott gap, with a small doping $x>0$ on the 
magnetic stripes (typically, $x\le 0.05$).

\subsection{Madelung Energies of Stripes}

We will assume for simplicity that all stripes, both magnetic and charged, are
an even number of cells wide.  This means that only a relatively small number
of stripe configurations are involved in the doping range of interest.  For
instance, labelling the stripe configuration by $m,n$, where $m$ is the width of
a magnetic stripe and $n$ the width of a charged stripe, we will explore in
detail the pure phases $m,n$ = 6,2 ($x=x_0/4\simeq 0.0625$, if $x_0\simeq 
0.25$), 4,2 ($x=x_0/3\simeq 0.0833$), 2,2 ($x=x_0/2\simeq 0.125$ -- the 1/8
phase), 2,4 ($x=2x_0/3\simeq 0.167$), and 2,6 ($x=3x_0/4\simeq 0.1875$).
Intermediate dopings would correspond to mixed phases.  For each of these 
phases, we assume that there can be different dopings on each row; by symmetry,
there can be $(m+n)/2$ inequivalent rows for the m,n-phase.

In the presence of charging, it is the electrochemical potential $\mu_e$ and
not the chemical potential, $\mu$, which is constant.  For electrons, 
$\mu_e=\mu -eV$, where $V$ is the electrical potential.  Given the average hole
density on each row, $V$ can be calculated as a Madelung sum.  
For each configuration, the Madelung sum can be calculated for each row.  
Actually, since the overall chemical potential must be adjusted to fix the
total hole density, all that need be calculated is the difference in Madelung
potential between the different rows.  This is calculated by assuming a pure
Coulomb interaction, screened by a static dielectric constant, $\epsilon$.  The
on-site term is neglected, having already been included as $U$.  

The various Madelung constants can be expressed as follows.  For the (6,2)
stripe, label the rows 1,2,3,4, with 4 = the charged stripe, and 1 (3) = the
magnetic rows farthest from (nearest to) row 4.  Let $V_i$ be the Madelung
potential for the $i$th row, $\tilde V_i=(V_i-V_1)x_0$, $x_i$ the hole doping of
the $i$th row, and $\tilde x_i=(x_i-x_1)/x_0$.  Then 
\begin{equation}
\tilde V_i=V_0\sum_jK_{ij}\tilde x_j,
\label{eq:11}
\end{equation}
where the $K^m$ matrices have been calculated numerically, with results listed 
in Table I, for the cases (m,2), $m$ = 2,4,6.  The constant $V_0=2x_0e^2/
(\epsilon a)=0.914eV/\epsilon$, for $x_0=0.25$.

\begin{tabular}{||c|c|c|c||}        
\multicolumn{2}{c}{{\bf Table I: Madelung Matrices}}\\ 
            \hline\hline
$K^m_{ij}$ & $j$ = 2&3&4 \\  
    \hline\hline
$K^6_{2j}=$ &-0.4110&0.5365&0.347          \\     \hline       
$K^6_{3j}=$ &-0.3466&0.4721&1.230          \\     \hline       
$K^6_{4j}=$ &-0.8831&0.8831&1.702          \\     \hline       
$K^4_{2j}=$ &-0.3082&0.6951&          \\     \hline       
$K^4_{3j}=$ &0.&1.082&          \\     \hline       
$K^2_{2j}=$ &0.1256&&          \\     \hline       
\end{tabular}
\vskip 0.2in

The stripe phase is stable only if the dielectric constant is large 
enough: 
recent calculations\cite{VBBK} suggest $\epsilon >5$ is sufficient.  The large 
static dielectric constant of the cuprates, $\sim 40-80$\cite{diel}, is a sign 
of strong electron-phonon coupling.  This large coupling makes it difficult to 
accurately estimate the strength of the Coulomb interaction.  The d.c. 
dielectric constant will be anisotropic and, most probably, dispersive, on the 
length scale of the stripes.  Since interlayer contributions to screening can be
important (e.g., apical oxygens, bilayer coupling), this is one parameter which
could easily have a strong material dependence.

While the above procedure should approximately capture the long range part of 
the Coulomb interaction, it will likely overestimate the hole-hole repulsion for
nearest neighbors.  This can be thought of in terms of a correlation hole having
two 
components.  First, we are assuming that a hole on a given site interacts with
a fractional hole (the average doping) on all other sites.  Clearly, part of 
the hole population on the nearest neighbor sites is actually generated by
the hopping of the given hole, hence should not be counted in the Madelung sum.
Moreover, there is likely to be a real correlation hole, as neighboring charges
readjust to avoid the given hole.  However, these terms are associated with 
CDW formation, which will not be dealt with explicitly here.

\section{Results}

\subsection{Absence of Coulomb Interaction}

\begin{figure}
\leavevmode
   \epsfxsize=0.33\textwidth\epsfbox{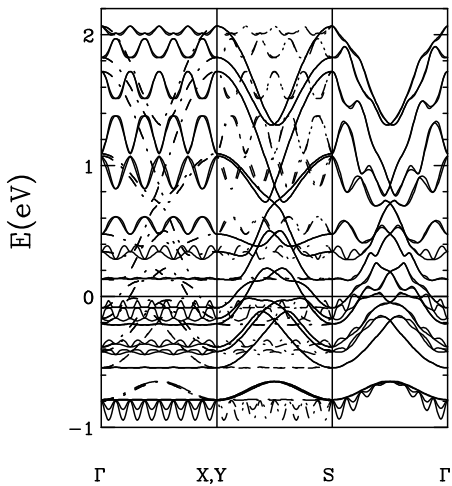}
\vskip0.5cm 
\caption{Dispersion of 2,6 structure. Solid lines $\Gamma\rightarrow X
\rightarrow S\rightarrow\Gamma$; dotdashed lines $\Gamma\rightarrow Y
\rightarrow S$. Here, $Y$ is along the stripes, $X$ is across them.}
\label{fig:11}
\end{figure}

Figure~\ref{fig:11} illustrates the band dispersion for a 2,6 structure ($x=
0.1875$), in the absence of long-range Coulomb effects.  The hole doping on each
layer is self-consistently adjusted to allow for inter-row hopping processes, 
and the Fermi level is adjusted to account for the overall doping.  In the 
absence of long-range Coulomb effects, the doping is close to the nominal 
values.  Numerical results will be discussed in the following subsection, which
will show how they are modified by Coulomb interaction.  

\begin{figure}
\leavevmode
   \epsfxsize=0.33\textwidth\epsfbox{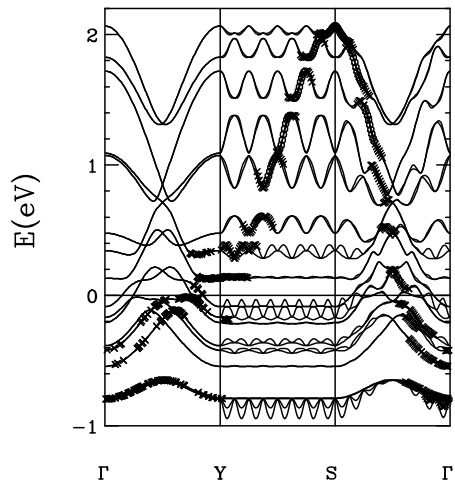}
\vskip0.5cm 
\caption{Dispersion of 2,6 structure, but with structure factors.}
\label{fig:12}
\end{figure}
\begin{figure}
\leavevmode
   \epsfxsize=0.33\textwidth\epsfbox{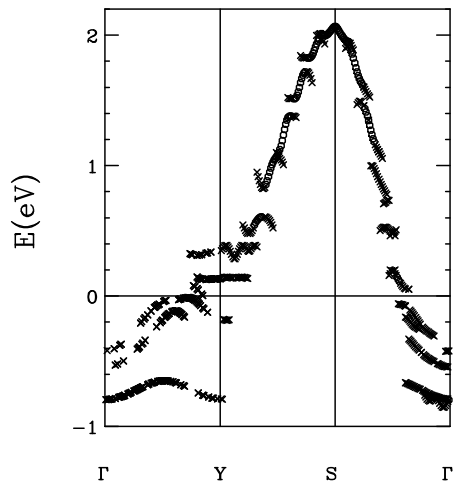}
\vskip0.5cm 
\caption{Dispersion of 2,6 structure, parallel to the stripes.}
\label{fig:14}
\end{figure}

\begin{figure}
\leavevmode
   \epsfxsize=0.33\textwidth\epsfbox{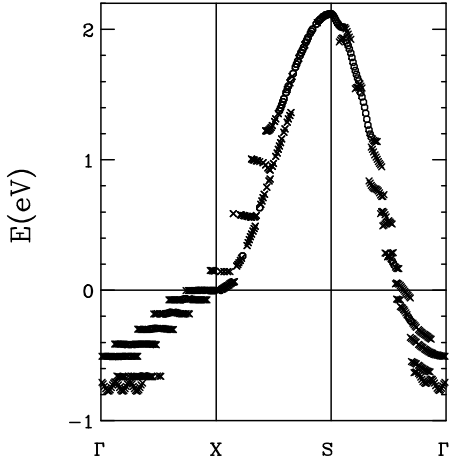}
\vskip0.5cm 
\caption{Dispersion of 2,6 structure, transverse to the stripes.}
\label{fig:13}
\end{figure}

The large number of bands is rather deceptive.  It is equal to the number of
Cu atoms in the large unit cell, doubled since the up and down spin 
bands are not degenerate.  There would be the same number of bands {\it even
if there were no stripes}.  But in this case, only one band would satisfy
Bloch's theorem.  This band can be determined by looking at the structure
factor -- the overlap of the
corresponding wave functions with $e^{i\vec k\cdot\vec r}$.  Similarly, when
stripes are present, the same structure factor determines which bands will be
seen by photoemission. This is illustrated in Fig.~\ref{fig:12}, where the
circles indicate a weight greater than 0.5, and the $\times$'s a weight between
0.5 and 0.1.  For greater clarity, only the dispersions along $\Gamma\rightarrow
Y\rightarrow S$ are shown.  The dispersions along $Y$ (parallel to the stripes) 
and $X$ (transverse to the stripes) are shown in Figs.~\ref{fig:14} and 
\ref{fig:13}, respectively.  The resulting weights reveal a simple result: the
envelope of the bands is approximately a superposition of the two limiting 
bands, at half filling and at optimal doping, with considerable fine structure
associated with minigaps.   

In the presence of stripes, the dispersion should be quasi one-dimensional.
This is clearly seen in Fig.~\ref{fig:11}, where the dispersion along $\Gamma
\rightarrow Y$ ($\Gamma\rightarrow X$) closely resembles that along $X
\rightarrow S$ ($Y\rightarrow S$).  However, with the structure factors 
included, the dispersions are quite distinct.  Nevertheless, the minigaps are
most prominent in the dispersions perpendicular to the stripes, $\Gamma
\rightarrow X$, (Fig.~\ref{fig:14}), and $Y\rightarrow S$, (Fig.~\ref{fig:13}).

\begin{figure}
\leavevmode
   \epsfxsize=0.33\textwidth\epsfbox{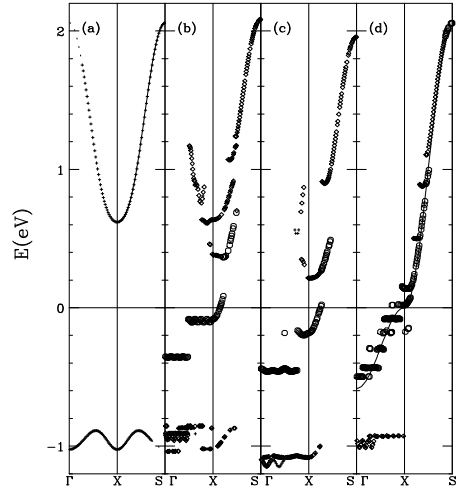}
\vskip0.5cm 
\caption{Total dispersion from $\Gamma\rightarrow S$ for dopings: $x$ = 0 (a),
0.0625 (b), 0.125 (c), 0.1875 (d), and 0.25 (solid line in d).  Data in (a) were
shifted upward by 0.16eV.}
\label{fig:21}
\end{figure}

Fig.~\ref{fig:21} shows the doping dependence of the dispersion, illustrating
the `projected' dispersion of the charged and magnetic stripes.  The projected
charge stripe is found by plotting those points where the wave function spends
$\ge 80\%$ of its time on the charged layers.  The curves show the dispersions
for a series of dopings, from $\Gamma\rightarrow X\rightarrow S\equiv 
(\pi ,\pi )$.  While the fine structure (minigaps) is
strongly doping dependent, {\it the overall dispersion is not}, and is 
essentially identical to the dispersion of the uniform end phases.  This is
exactly what would be expected for {\it macroscopic} phase separation, even 
though at crossover the charge stripe is only two cells wide.  

The figure shows that the dispersion is largely a superposition of two Fermi
surfaces: one for the insulating magnetic stripes, one for the charged stripes:
the small $+$'s (large circles) indicate $\ge 80\%$ of the wave function is on 
the magnetic (charged) stripe; the small diamonds indicate a mixture of both.
Note that there is strong overlap in the region of the upper Hubbard band, while
the magnetic lower Hubbard band (LHB) remains well defined at all dopings, and 
the charged stripes fill in the gap as doping increases.

It should be noted that once the charged stripes are reduced to two cells wide,
at $x$=0.125, the dispersion remains nearly unchanged as the doping is further
reduced (e.g., at x=0.0625).  Hence, an important aspect to understanding the
strongly underdoped stripe phases will be to develop a good model for these 
limiting, two-cell stripes.  As discussed below, there is an analogous magnetic 
stripe beyond the percolation crossover, which can be modeled as a two-leg 
ladder.

\subsection{Coulomb Interaction}

\begin{figure}
\leavevmode
   \epsfxsize=0.33\textwidth\epsfbox{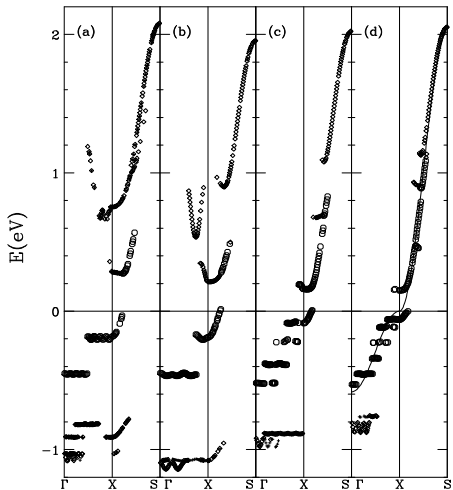}
\vskip0.5cm 
\caption{Total dispersion corrected for charging ($\epsilon =15$) from 
$\Gamma\rightarrow S$ for dopings: $x$ = 0.0625 (a), 0.125 (b), 0.167 (c), 
0.1875 (d), and 0.25 (solid line in d).}
\label{fig:21b}
\end{figure}

Inclusion of Coulomb interaction leads to fairly modest changes in the 
dispersion.  Figure~\ref{fig:21}, with no Coulomb effects, should be compared to
Fig.~\ref{fig:21b}, with moderate screening $\epsilon =15$.  Careful inspection
reveals that the charged bands are shifted to lower energy with respect to the
magnetic layers, so that the lower Hubbard band is more fully hybridized with
the charged layers.  With reduced screening ($\epsilon =5$) the bottom of the
charged band actually falls below the magnetic lower Hubbard band.
The layers near the Fermi level remain predominantly
associated with the charged layers, so we may still loosely speak of charged
bands and magnetic bands.  Note that in every case, the Fermi level lies within
the minigap closest to the Van Hove singularity.  This provides a {\it new
mechanism} for the opening of the pseudogap, as will be discussed further below.

\begin{figure}
\leavevmode
   \epsfxsize=0.33\textwidth\epsfbox{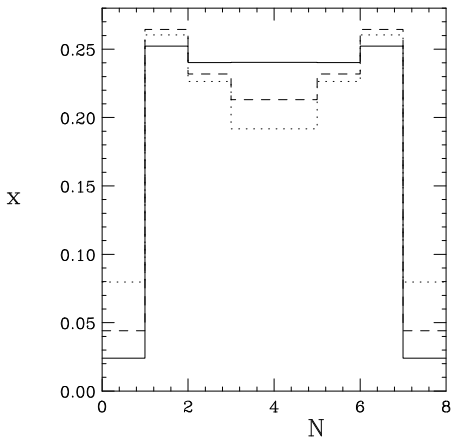}
\vskip0.5cm 
\caption{Hole distribution on rows (labelled by $N$) of the 2,6 structure, 
for $\epsilon$ = $\infty$ (i.e., no Coulomb interaction -- solid line), 15
(dashed line), or 5 (dotted line).}
\label{fig:17}
\end{figure}

Even in the absence of Coulomb interaction, the carrier density in a given row
deviates somewhat from the free energy minima -- here taken as $x$= 0, 0.25 --
due to the finite hopping probability.  For the 2,6 structure,
Figs.~\ref{fig:11}-\ref{fig:14}, the magnetic layers have $x$=0.025, and for the
charged layers, moving 
away from the magnetic layer, the hole doping is 0.25, 0.24, and 0.24.  Adding 
the Madelung potential raises the energy of the hole-doped stripes, and requires
a shift of the Fermi energy to compensate.  However, since the magnetic stripes 
are gapped, this shift makes little difference to the hole population on these
stripes, the corresponding layer populations being 0.026, 0.255, 0.25, and 0.22,
for $\epsilon =15$, Fig.~\ref{fig:21b}.  For larger Coulomb interaction,
the deviation becomes greater, Fig.~\ref{fig:17}.  The data display an 
interesting evolution: superimposed on a trend toward greater homogeneity, there
is also a tendency to evolve into a (2,2) state.  This can be understood from
Table I: the Coulomb effects are smallest for this state, since the phase 
separation is restricted to the finest scale.

This result is of potential relevance for LSCO: it is found experimentally that
the incommensurability saturates near 1/8 doping -- here the crossover where the
(2,2) phase is stable.  The saturation could simply mean that for LSCO, the
Coulomb effects are large enough that the system locks into the (2,2) phase for
all higher dopings. 

There is a striking asymmetry about 1/8 doping: in the (6,2) phase, Coulomb
interaction makes very little difference.  This is because of the sharp cusp
instability at half filling, which keeps the hole doping fixed near zero in
the magnetic stripes, whereas the shallower potential minimum near optimal
doping allows more substantial density fluctuations.

These results will be discussed further in a later section.  Two points are 
worth
mentioning: first, the incommensurability saturation has so far only been
observed in LSCO; and secondly, LSCO closely resembles the other cuprates in the
doping range up to 1/8, but for higher doping, T$_c$ saturates at a much lower
value.  

\subsection{Minigaps}

Figure \ref{fig:21b} shows the evolution of the minigaps with doping.  A simple
model provides a semiquantitative explanation of these results.  The
dispersion along $(0,0)\rightarrow (\pi ,0)$ is discretized into $n$ levels
for $n$-Cu wide charge stripes.  This bandwidth is $4(t+2t^{\prime})\simeq
584meV$.  If the minibands are equally spaced, the average gap should be
$584/(n-1)$ meV.  Actually, the net bandwidth changes some with doping, so a
better formula is 
\begin{equation}
\Delta_{av}={584 meV\over n}
\label{eq:11e}
\end{equation}
= 292 ($n$=2), 146 (4), or 97 (6) meV, to be
compared with average values (Fig. \ref{fig:21b}) of 260, 147, and 94 meV, 
respectively.  For the dispersion along $(0,\pi )\rightarrow (\pi ,\pi )$, the
same bands are present, but shifted by the dispersion along $Y$, and with
total bandwidth $4(t-2t^{\prime})$.

For fluctuations in the stripe spacing, there will be a tendency to average
over the various dispersions in Fig. \ref{fig:21b}.  This will tend to wash out
most of the minigaps, since they are shifted in energy as the stripe width
changes.  However, since there is always one gap present near the Fermi level,
this gap should survive averaging.  For the uniform stripe phases of 
Fig. \ref{fig:21b}, this 'pseudogap', or distance between the Fermi level and 
the nearest $(\pi ,0)$ minigap, follows the same scaling as Eq. \ref{eq:11e}, 
$\Delta_p=364/n$ meV.

\subsection{Ferromagnetic Stripes}

\begin{figure}
\leavevmode
   \epsfxsize=0.33\textwidth\epsfbox{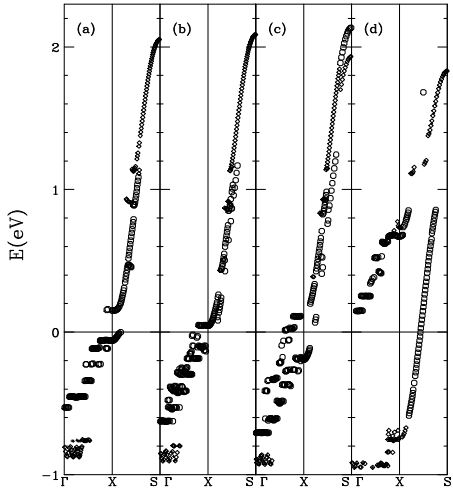}
\vskip0.5cm 
\caption{Dispersion along $X$ for a ferromagnetic instability on the charged
stripes, for $x=0.1875$ (2,6), and different degrees of magnetization $M$, as 
discussed in the text.}
\label{fig:18}
\end{figure}

Figure \ref{fig:21} shows that, beyond the percolation
crossover a clear remnant of the bulk VHS is visible in the striped phase
dispersion.  In Figure \ref{fig:18}, it can be seen that splitting this
VHS degeneracy produces a clear pseudogap-like splitting of the dispersion near
$(\pi ,0)$.  It is this lowering of a large density of states that has been
postulated to stabilize the charged stripes, and the figure clearly shows that
the mechanism remains active even in the striped phase.

\begin{figure}
\leavevmode
   \epsfxsize=0.33\textwidth\epsfbox{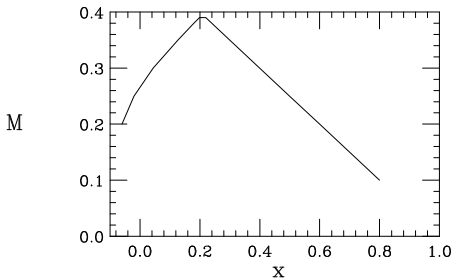}
\vskip0.5cm 
\caption{Doping dependence of the magnetization $M$ for a ferromagnetic
instability.}
\label{fig:18a}
\end{figure}

For the calculations in the figure, it was assumed that the antiferromagnetic
phase is stable only up to a doping $x_0/3$, while for larger doping a
ferromagnetic instability wins out.  The ferromagnetic dispersion is also given 
by Eqs.~\ref{eq:A0}-\ref{eq:A2}, but with $q=(0,0)$ instead of $Q=(\pi ,\pi )$. 
For the same value of $U$, the equilibrium $M$ has the form shown in 
Fig.~\ref{fig:18a}, which was approximated by $M=0.4-0.5|x-0.2|$.

It should be noted that the doping dependence depends sensitively on the
choice of parameters; these values are taken for illustrative purposes only.
Figure \ref{fig:18}d shows the dispersion of the self-consistent solution with
the full $M(x)$, while the other frames show a reduced $M$ of 1/5 (c), 1/10 (b),
or 0 (a).  Since the parameters were chosen 
to have the VHS in frame (a) centered on the Fermi level, the pseudogap opens
approximately symmetrically about the Fermi level.

This should not be taken as evidence that the charged stripes really are
ferromagnetic, only as an example of yet another kind of instability that is
driven by the VHS.  The figure illustrates that one can distinguish different
instabilities, but one must carefully analyze secondary characteristics, since
the opening of the pseudogap near $(\pi ,0)$ is common to a variety of
instabilities.  In the present instance, a ferromagnetic instability does not 
double the unit cell, so the ghost dispersion beyond $(\pi ,0)$ is absent,
in contrast to experiment (see Fig. \ref{fig:B3} below).  Moreover, the
splitting of the spin up and spin down bands should lead to extra structure
most clearly seen (below the Fermi level) near $\Gamma$, which is not found
experimentally.

\section{Comparison to Experiment}

\subsection{Photoemission in LSCO}

The doping dependence of the photoemission spectra in LSCO\cite{Ino} is 
strikingly different from that in BSCCO\cite{Bdope}.  In this section, it
will be shown that {\it both} spectra can be interpreted in terms of stripe
phases, with stronger fluctuation effects in BSCCO.

\begin{figure}
\leavevmode
   \epsfxsize=0.33\textwidth\epsfbox{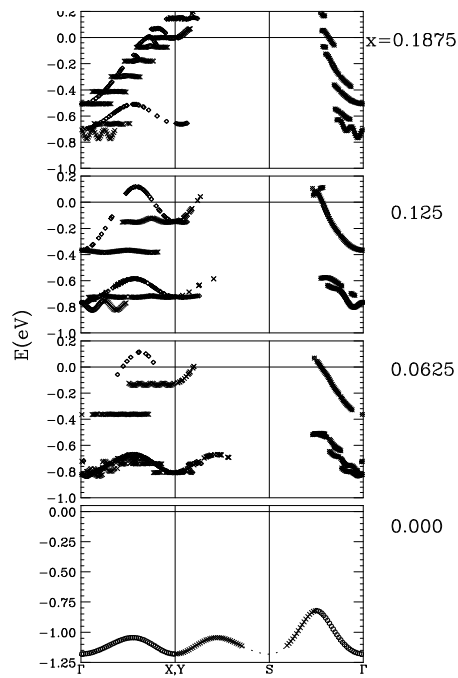}
\vskip0.5cm 
\caption{Superposition of $X$ and $Y$ dispersions for, from bottom to top, $x$ =
0, 0.0625 (6,2 structure), 0.125 (2,2), and 0.1875 (2,6).}
\label{fig:15}
\end{figure}

Since no-one has yet
succeeded in making a single-domain sample, the photoemission should best be
compared with a superposition of the $X$ and $Y$ dispersions.  
Figure~\ref{fig:15} illustrates how the pattern changes
with doping.  The following features should be noted: (1) there is always a
prominent flat band near $(\pi ,0)$, which tends to shift further below the
Fermi level with increased underdoping.  (2) By construction, the dispersion at 
half filling matches that found in SCOC.  (3) The evolution with doping is not 
smooth, but crosses over from the SCOC band near half filling to a more metallic
band near the Fermi level at higher dopings.  The overall doping is quite 
similar to the experimental results of Ino, et al.\cite{Ino} (see particularly 
their Fig. 3), confirming the suggestion that stripes are better developed in
LSCO than in BSCCO.

Figure \ref{fig:31} compares the $(\pi ,0)$ photoemission peak positions for
LSCO\cite{Ino} with the present calculations.  Results for BSCCO\cite{pdh2} are 
also shown; these will be discussed in the following subsection.  In LSCO, 
there are two main features: one ($\times$'s) is near -0.6eV, with a dispersion 
similar to that in the magnetic insulator SCOC, and with a nearly 
doping-independent binding energy. The second feature (open circles) is a gap
close to $E_F$ with larger doping dependence.  Qualitatively, these features are
similar to the hump and peak features in BSCCO, but with larger binding
energies.  These two features can be correlated with two prominent gap-like
features in the calculations: the magnetic gap associated with the lower
Hubbard band on the magnetic stripes, and the charge stripe gap, associated with
the miniband closest to the Fermi level.  The calculated gaps are larger, since
the energy scale has been chosen to agree with the magnetic gap in SCOC, 
yielding a value -1.2eV at half filling, but the overall doping dependences are 
quite similar to LSCO.  This similarity is brought out most clearly by plotting 
the calculated gap values divided by two (diamonds and suns).  

\subsection{Photoemission in BSCCO}

\subsubsection{Below T$_c$}

The photoemission in BSCCO is strikingly different from the above results, yet
also provides evidence for stripes, but of a more fluctuating form.  In BSCCO,
there is a remarkable evolution of the photoemission with temperature, 
particularly on passing through T$_c$.  Above T$_c$, the spectra are very broad,
with a single broad peak near $\pi ,0)$ representing the normal-state 
pseudogap.  Below T$_c$, the spectra sharpen and split into two features, 
commonly referred to as a `peak' near $E_F$, with a `hump' at lower energies,
close to the normal state pseudogap; between the peak and hump, there is a clear
`dip' in intensity, below the level in the normal state.  Recently, systematic 
studies of these features in both tunneling\cite{pdh1} and 
photoemission\cite{pdh2} were presented.  Most strikingly, photoemission finds 
these two peaks in the same direction of $k-$space, a feature which is very
suggestive of phase separation.

Here, it will be assumed that the photoemission is dominated by stripe effects,
and the main role of superconductivity is to suppress fluctuations.  (The clear
sharpening of the spectra below T$_c$, even in a range away from any gaps, is 
demonstrated in Ref.~\onlinecite{sharp}.)  The 
analysis will be in two parts.  First, the low-T spectra will be compared with
those of LSCO.  Then the role of fluctuations in producing the high-T smeared
spectra will be discussed.

\begin{figure}
\leavevmode
   \epsfxsize=0.33\textwidth\epsfbox{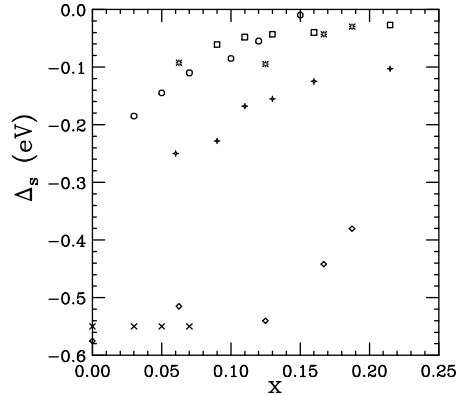}
\vskip0.5cm 
\caption{Pseudogaps at $(\pi ,0)$ in LSCO\protect\cite{Ino} ($\times$'s, open 
circles) and BSCCO\protect\cite{pdh2} ($+$'s = hump, squares = peak) compared 
to calculated Mott gap (diamonds) and minigap (suns); the calculated values are
reduced by a factor of two. At the highest doping, the Mott gap does not show
much intensity near $(\pi ,0)$; what is plotted is energy of the corresponding 
dispersion nearest to $(\pi ,0)$.}
\label{fig:31}
\end{figure}

In Figure \ref{fig:31} the low-temperature photoemission peaks\cite{pdh2} of
BSCCO are compared to those of LSCO.  The `peaks' (squares) are in reasonable 
agreement with the near-$E_F$ pseudogap in LSCO (circles), and with the 
calculated minigaps.  On the other hand, the `humps' ($+$'s) are considerably 
closer to the Fermi level than the magnetic stripe feature in LSCO ($\times$'s);
compared to theory, the overall offset is different, but the doping dependence
is similar.  Nevertheless, identification of the hump 
with the antiferromagnetic Mott gap feature is compelling.  Laughlin\cite{Bdope}
clearly showed that the photoemission data evolve with doping to match the SCOC
spectrum at half filling.  A detailed fit shows that this only works if the
dispersion at $(\pi /2,\pi /2)$ touches the Fermi level -- unlike SCOC, where
there is a $\sim 0.8$eV gap or LSCO with a $\sim 0.3$eV gap.  In Appendix B it 
is shown that the complete doping dependence of the hump is well described by 
simply doping into the lower Hubbard band of the antiferromagnet.

There is a clear progression in the magnetic stripe separation from the Fermi
level, from SCOC (and the present magnetic stripe calculation) to LSCO to BSCCO.
It is likely that this is due to stripe fluctuation effects, since as the band 
moves closer to $E_F$, there will be progressively more holes on the magnetic 
stripes.  Further evidence for this interpretation lies in the high-T BSCCO 
spectra, where the peak and hump collapse into a single feature, which continues
to resemble the magnetic dispersion (Appendix B) and is even closer to $E_F$ 
than the hump.  The conclusion that fluctuations are strongest in BSCCO is 
consistent with the fact that incommensurate magnetic modulations have not yet 
been clearly seen in BSCCO.  

\subsubsection{Above T$_c$}

In BSCCO, there is a sudden change of the photoemission spectrum at T$_c$: a
single broad feature above $T_c$ splits into a peak-dip-hump structure below 
$T_c$.  This is here assumed to be mainly a fluctuation effect: above $T_c$, 
both features are assumed to be present, but the line broadening is so large
that they strongly overlap.  Below T$_c$, fluctuations are greatly suppressed, 
and the linewidth broadening $\Gamma$ is reduced by over an order of 
magnitude\cite{eesup}, 
so a truer picture of the spectra is obtained.  Superconductivity will also
renormalize the gap on the charged stripes (as discussed for a related model in
Ref.~\onlinecite{MKK}), but this will be a secondary effect.

In LSCO, the photoemission spectra were observed\cite{Ino}
only in the superconducting state, due to surface degradation at higher 
temperatures.  However, the stripes are clearly better defined in LSCO (the
two gap features are more clearly separated) even though T$_c$ is considerably
lower, so it is quite possible that the split spectral peaks persist above 
T$_c$.

Ref.~\onlinecite{MKK} (Fig. 19) showed that the changes in BSCCO at T$_c$ can 
be interpreted in terms of enhanced fluctuations in the normal state; a similar
calculation was presented by Chubukov and Morr\cite{ChuM}.  
To explain the observed BSCCO photoemission, it is assumed that the local hole
density is inhomogeneous, and the photoemission can be described as a 
superposition of the spectra of different densities.  The spectral
broadening in the superconducting state is taken to have the normal-state value
for energies greater than twice the superconducting gap, $\Delta_s$, but to have
a (5$\times$) smaller value at lower energies.  This spectral sharpening can
produce a double-peaked photoemission spectrum, even if the density distribution
is single peaked.  A detailed comparison with experiment, Fig. 4a of 
Ref.~\onlinecite{pdh2}, will require a more detailed model of the fluctuations.

There is an alternative class of models for the
peak-dip-hump structure.  In this picture, the dispersion is split because the
holes strongly interact with some bosonic excitation.  
Abanov and Chubukov\cite{AChu} have related the dip position to the resonance 
peak seen in neutron scattering.
It is tempting to speculate that both views are approximately correct, and that 
the bosonic excitations originate from stripe fluctuations.

\section{Duality Crossover: A Percolation QCP}

There is considerable evidence for a quantum critical point (QCP) in the 
cuprates\cite{Aep}.  However, the QCP is at too high a doping to be a 
conventional N\'eel QCP, which should fall at a doping near $x\simeq 
0.02$, where $T_N\rightarrow 0$.  Moreover, in undoped La$_2$CuO$_4$ and other
spin-1/2 antiferromagnets, no evidence is found for a high-T crossover to a
quantum critical state\cite{QC}.
The QCP also lies significantly below optimal doping\cite{Schma}.  The QCP has 
also been associated with the termination of stripe ordering\cite{QCP}, but 
again, the doping seems wrong: stripes persist well above optimal doping in 
Nd-substituted LSCO, as does the pseudogap.  The
evidence seems to point to the 1/8 anomaly playing a role: Hunt, et
al.\cite{Imai} find a significant crossover, which they suggest is the end of
the stripe regime, at $x\simeq 0.125$, and they further suggest that the
high-field metal-insulator transition\cite{Boeb} takes place at this point,
rather than at optimal doping, as reported earlier.

The present model suggests an attractive alternative interpretation for the QCP,
as a {\it duality crossover} of the striped phase.  (For other patterns of
nanoscale phase separation, such as islands, this would correspond to a
percolation crossover, but for 1D stripes, there is no true percolation.)  At 
this doping, both
magnetic and charged stripes have their minimum width, two cells.  For lower
doping, the magnetic cells widen, reducing coupling between the charged stripes,
while for higher doping, it is the magnetic stripes that decouple.  
Correspondingly, below $x=0.125$, there is a charge gap, leading to high-field
localization, while above 0.125 there is a spin gap.  

Note that this is not a conventional QCP, where there is an abrupt change of
groundstate at T=0.  For instance, for the 2D Hubbard model, on one side of
the QCP there is believed to be a renormalized classical regime, with finite 
N\'eel order, and on the other side a quantum disordered regime, with a spin 
zero groundstate and a finite gap to the lowest triplet excitation.  Instead,
in the stripe crossover, magnetic layers persist on both sides of the duality
crossover.  

While both phases persist across the duality point, there is a crossover in the
nature of the majority phase (magnetic or charged), and consequently the 
properties of the minority phase are strongly modified.  Because of this, 
modifications need not occur exactly at the phase boundary.  Instead, there
strong changes in properties may be observed at any of a series of {\it magic
dopings}, corresponding to the commensurate stripe phases discussed above.  

For instance, the localization transition in LSCO occurs not exactly
at x=0.125, but closer to 0.17.  This could be understood as evidence that 
two-leg charged stripes are always localized, but 4-leg stripes are not.  For
a uniform stripe phase, the two-leg stripes would just disappear at $x=2x_c/3
\simeq 0.167$.  However, the localization behavior does not seem to be
universal: in heavily underdoped YBCO, the resistivity of the normal state
saturates at low temperatures, suggestive of a metallic state\cite{SeAn}.

Again, the horizontal-vertical stripes are replaced by diagonal stripes at a
doping 0.058, where the superconducting transition terminates.  This is close
to the doping of the (6,2) phase.  (Note that the precise value of the magic
dopings depends on $x_0$; the pure (6,2) phase would fall at x=0.058 if $x_0=
0.23$.)

Similarly, while there is not a conventional quantum disordered 
regime, a spin gap can still arise when the magnetic phase is the minority
phase, although not necessarily at x=0.125.  Indeed, in LSCO at x=0.12 there
is well-defined long-range (incommensurate) N\'eel order\cite{Suz}, with
$T_N\simeq T_c$, the superconducting transition temperature.  A simple
interpretation of this result is that superconductivity is predominantly
associated with the charged stripes, and the superconducting transition 
enhances charge phase stiffness, reducing the fluctuations of the charged
stripes which were suppressing magnetic order on the magnetic stripes.  At 
higher doping, the magnetic stripes are gradually spread apart, behaving more
like two-leg ladders, and if the inter-ladder coupling becomes 
sufficiently weak, a spin gap can open up on each ladder.  This is discussed 
further in Subsection VII.A.

Note that, while a duality crossover is a generic feature in a stripe model
based on {\it phase separation}, there are alternative stripe models\cite{Cod}
wherein the charged stripes are merely domain walls of antiferromagnetic 
domains, and a charge-stripe dominated regime would be meaningless.

\section{Isolated Magnetic Stripe}

In the stripe phase away from the duality 
crossover, the minority phase is present in the form of domain walls between
domains of the majority phase.  An important aspect of stripe phase theory is 
the development of a microscopic model for these domain walls.  A number of
groups have suggested a connection between magnetic stripes and even-leg 
ladders.  Here, the magnetic domain walls in the higher-doping regime are 
modelled as two-leg ladders, which develop a spin gap as they move further
apart, with reduced interladder coupling. 

\subsection{Spin Gap}

In a stripe model, the magnetic neutron scattering near $(\pi ,\pi )$ should be
reflective of the properties of the magnetic stripes.  For LSCO,
the incommensurability has been discussed above, Fig.~\ref{fig:17}, and is 
further discussed in the following subsection.  In YBCO, incommensurability 
has only been resolved at one doping\cite{M1,Arai}, but the doping dependence of
the peak width is consistent with a similar underlying, but unresolved 
incommensurability\cite{B1}.  In YBCO, the stripe model can also explain the 
doping dependence of the intensity of the magnetic neutron scattering near 
$(\pi ,\pi )$, as well as the opening of a {\it spin gap}.

\begin{figure}
\leavevmode
   \epsfxsize=0.33\textwidth\epsfbox{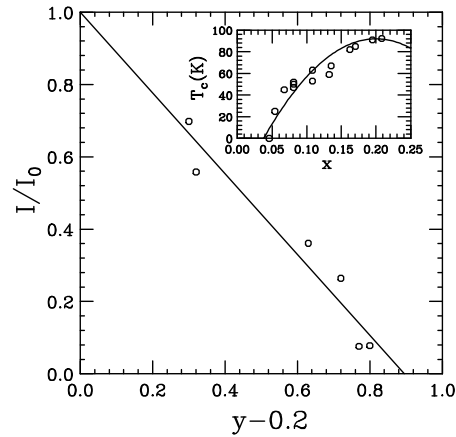}
\vskip0.5cm 
\caption{Intensity of magnetic scattering vs. doping for YBCO.  Circles = data 
of Ref.~\protect\cite{R-M}; line = expected result for stripe model.}
\label{fig:7b}
\end{figure}

The doping dependence of the net intensity of the magnetic neutron scattering 
should reflect the relative density of magneticb stripes.  For YBCO$_{6+y}$, the
intensity was numerically integrated from Fig. 2 of Ref.~\onlinecite{R-M}, and 
the result plotted in Fig.~\ref{fig:7b}.  While the 
relation between $y$ and hole doping $x$ in YBCO is not completely settled, 
the straight line illustrates a modified Tokura\cite{Tok} expression, with
the doping of the planes starting at $y=0.2$, and varying linearly with $y$. 
The results are consistent with the picture that all magnetic scattering is 
associated with the magnetic stripes, and the stripe phase would terminate at an
(inaccessible) doping $y=1.095$.  This would place the percolation crossover
at $y\sim 0.65$, close to the plateau regime.  Since the plateau has been 
interpreted as a 1/8 effect\cite{Tal2,ANO}, this suggests that the plateau 
doping is $\sim 0.125$.  This fixes the constant of proportionality:
$x=0.27(y-0.2)$, so the charged stripe doping, corresponding to $y=
1.095$, would be $\sim 0.25$, in excellent agreement with our other estimates.
At optimal doping, $y\sim 0.925$, the hole doping would be $\sim 0.2$.  These
estimates are also consistent with Tokura, et al.\cite{Tok}, who found $x=
0.125$ for $y=0.75$, $x=0.25$ for $y=1$,and $x=.21$ for optimal doping.
The inset to the figure shows that T$_c$(x) follows the familiar parabolic
form\cite{Tal}, 
\begin{equation}
{T_c\over T_{c,max}}=1-({x-x_m\over x_w})^2,
\label{eq:1e}
\end{equation}
with $T_{c,max}=92K$, $x_m=0.2$, and $x_w=0.16$.
Note that the dip in $T_c$ near the 60K plateau is close to $x=1/8$. 

The falloff of intensity is likely to be even steeper than illustrated in 
Fig.~\ref{fig:7b}, since the data were presented only up to about 60meV, while
at the lower doping levels, there is considerable intensity at higher
frequencies.

\begin{figure}
\leavevmode
   \epsfxsize=0.33\textwidth\epsfbox{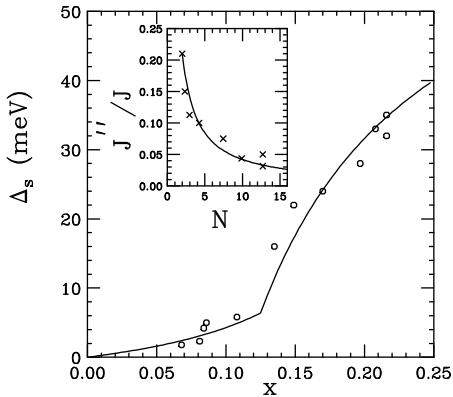}
\vskip0.5cm 
\caption{Spin gap $\Delta_s$ vs. doping for YBCO.  Circles = data of 
Ref.~\protect\cite{R-M}; line = theory, assuming solid line from inset. 
Inset: interladder exchange vs. hole-doped stripe width.}
\label{fig:7}
\end{figure}

The spectrum of the excitations near $(\pi ,\pi )$ has a complicated evolution
with doping, and below the superconducting $T_c$, the intensity is suppressed
below a doping dependent energy, called the `spin gap'\cite{R-M}.  This gap is 
distinct from the pseudogap, and has a strikingly different doping dependence, 
Fig.~\ref{fig:7}.  A similar gap is seen in LSCO\cite{LSgap,LAM}, but so far 
only near $x=0.15$.  There it is found that the spin gap is isotropic\cite{LAM},
further evidence that it is distinct from superconductivity or the pseudogap.

The doping dependence of this spin gap in YBCO can be 
interpreted simply in terms of coupled magnetic ladders, Fig.~\ref{fig:7}.  
Below the 1/8 crossover, the 
magnetic stripe (ladder) width decreases smoothly with doping, while the 
interladder coupling is approximately constant, since the hole-doped stripe has 
fixed width.  Theoretically, the spin
gap is found to be (approximately) inversely proportional to the ladder
width\cite{RiD}, so in this regime the spin gap scales linearly with doping,
$\Delta_s=\beta J/M$, where $J$ is the exchange constant, $M$ the ladder
width, and $\beta$ a correction for interladder coupling, $\beta\simeq (1-
4J^{\prime\prime}/J)$, with $J^{\prime\prime}$ the exchange coupling between 
adjacent ladders\cite{Ric2}.  The solid line in Fig.~\ref{fig:7} corresponds to
$J^{\prime\prime}=0.21J$.
\par
Above the crossover, $x>x_0/2=0.125$, $M$ is fixed at 2 while $\beta$ increases 
with doping, since $J^{\prime\prime}$ decreases as the hole-doped stripes widen.
Since the Cu in the hole-doped stripes can be magnetized, the falloff should be
relatively slow.  Details are model sensitive, but qualitatively the observed
behavior is readily reproduced.  The curve in Fig.~\ref{fig:7} follows from
assuming a falloff $J^{\prime\prime}\sim N^{-1}$, where $N$ is the hole-doped
stripe width, inset to Fig.~\ref{fig:7}.  (It should be noted that the falloff
is sensitive to the hole-density $x_0$, here taken as 0.25.)
\par
The model predicts\cite{Ric2} that for an isolated stripe, the spin gap equals
$J/2$, at least when the exchange constant is the same on all rungs and links.
From Fig.~\ref{fig:7}, this implies a limiting value $J\sim 80meV$ at $x=0.25$,
considerably smaller than the $x=0$ value $J=130meV$.  Such a doping dependence 
for $J$ is not unexpected.  For simplicity, however, the model assumes a
constant value for $J$;  this value must be taken as $J=80meV$, to successfully
model the single stripe limit.  The value is less critical near zero doping,
where the gap is small.

In this model, the spin gap should already exist in the normal state.  The 
striking change observed at $T_c$ can be explained as a fluctuation effect,
similar to those seen in the BSCCO photoemission.
Strong fluctuations at high temperatures prevent any long-range
stripe order or true spin gap.  The superconducting transition leads to 
three-dimensional coherence, and hence greatly suppresses charge and spin
fluctuations in the stripes.  Hence, a long range spin gap can open on the
magnetic stripes below $T_c$.  Consistent with this interpretation, it should be
noted that the spectrum in the normal state in heavily doped YBCO has been 
interpreted\cite{Bourg,LAM} in terms of a formula derived for spin-1 
chains\cite{ReZ}, and hence expected to approximately hold for spin 1/2, 2-leg 
ladders.

\subsection{Incommensurability Saturation}

Inelastic magnetic neutron diffraction finds a saturation of the 
incommensurabilty in LSCO at approximately $x=1/8$.  Within the present 
framework, there are actually several possible explanations for the saturation.
One was discussed above, Fig. \ref{fig:17}: strong Coulomb interactions arrest
the phase separation at the (2,2) stripe, and higher doping causes these 
stripes to gradually fill in.

On this interpretation, the Coulomb effects are much stronger in LSCO than in
YBCO, and the isolated spin gap regime (right-hand side of Fig.~\ref{fig:7})
would not exist in LSCO.  The Coulomb effects would indeed be expected to be
stronger in LSCO, since interlayer screening is weaker.  In Nd-substituted
LSCO, due to the LTT phase structural distortions, stripes in alternate layers
are rotated by 90$^o$; a similar situation may arise in LSCO, perhaps due to
LTT fluctuations.  On the other hand, in YBCO the magnetic correlations have a
strong c-axis modulation, suggesting that stripes in both CuO$_2$ planes of a
bilayer run parallel, with the charged stripes offset laterally to provide
stronger interlayer screening.

However, there are other plausible explanations for incommensurability
saturation.  Even before the discovery of
stripes, it was found that LTO and LTT domains of fairly large size 
(producing distinguishible diffraction peaks) coexist near 1/8 doping in
LBCO.  It seems plausible that this is associated with a stripe
commensurability effect, similar to that found in the nickelates, and that a
similar effect arises, at least incipiently in LSCO.  In this case, the 
residual magnetic scattering would be due to regions that have not yet been 
doped beyond 1/8.  In the more highly doped domains, the magnetic stripes 
would have a spin gap: since the ground state of a two-leg ladder is a spin 
singlet, it does not contribute to the magnetic scattering.  A related problem 
has been studied by Kim, et al.\cite {Birg2}, who showed that in a random mix of
weakly coupled three-leg (magnetic) ladders and two-legged (spin-gapped, and 
hence non-magnetic) ladders, the magnetic incommensurability remains unchanged
from that of the pure array of three-legged ladders.

Even without commensurability effects, one would expect 1/8 lock-in over a
finite doping range, when the 2,2 stripes coexist with 2,4 stripes, which have
a well defined spin gap.  In this case, the magnetic 
incommensurability should be fixed at that for 1/8 throughout the coexistence
regime, but should disappear when a commensurate 2,4 phase is stable, at
$x=2x_0/3\simeq 0.17$.  In LSCO, the 1/8 stripes are actually 
found\cite{Tran,Yam} to persist up to $x\simeq 0.25$.  Hence the need to
postulate lock-in effects at 1/8 doping, the exact analog of the stability of 
the $x=1/3$ and 1/2 striped phases in 
nickelates.  In this case, the heavily doped phases would have no magnetic
scattering, while the 1/8 stripes would have a scattering of fixed 
incommensurability, but decreasing intensity and increasing width, as the
stripe domains shrink in size.  
The special stability of the 1/8 phase may be associated with the finite
residual exchange coupling across the two-cell-wide charged stripes, which is
responsible for the antiphase boundaries, and which may be lost in wider
charged stripes, or with the reduced Coulomb energy.

At this stage, there is not enough information to judge between the two models
for incommensurability saturation.  The former, strong Coulomb effect, has the
advantage that it could simultaneously explain why T$_c$ in LSCO is so low --
the local hole density is forced away from optimal.  However, there is
considerable evidence that stripe phase order is better developed in LSCO than
in other cuprates, and this could provide reason enough for a lower T$_c$.

In many ways, optimally doped LSCO resembles an underdoped YBCO.  We have
here suggested that this is because stripes and pseudogaps in both materials
persist up to $x\simeq 0.25$, whereas T$_c$ is optimal near 0.16 in LSCO, 0.2
in YBCO.  Sato, et al.\cite{SYNT} have recently provided additional evidence
that the pseudogap opens well above T$_c$ in optimally doped LSCO.

\section{Discussion}

\subsection{Improvements for the Stripe Model}

The present model of the stripe phase provides a significant advance over
earlier calculations.  In these 
early calculations, an external stripe potential was imposed, which could be
either periodic or random, and the rearrangement of holes was studied.  It was
found that the stripes produced minigaps, but that an average, smeared 
dispersion and Fermi surface could still be defined.  However, there was no
sign of the split dospersion (magnetic vs charged stripe) found in photoemission
experiments.  The 
present calculation replaces the external potential with a given, doping
dependent potential, and studies how holes redistribute in the presence of a
competition between that potential, which favors phase separation, and Coulomb
repulsion, which favors a uniform density distribution.  The striking result is
that the resulting dispersion resembles a weighted superposition of the 
dispersions of the two end phases, with the addition of some superlattice 
minigaps.  This is very encouraging, in providing an explanation for the
photoemission results.  Moreover, it shows that the mechanisms 
responsible for the special stability of the end phases can continue to operate
on these nanoscopic length scales.

A number of improvements still need to be made in the model.  The next step 
would be to make the calculation fully self-consistent, by
eliminating any assumed potential, and directly calculating and minimizing the
free energy of the striped phase.  Since the dispersion is not greatly changed,
it is unlikely that this additional step will greatly modify the present 
results.  The most likely change would be that the densities could adjust 
slightly to take advantage of the minigaps, better centering them at the
Fermi level.  This could lead to a more systematic growth of the pseudogap with
underdoping, since the minigaps are associated with the charged stripes, and
get larger as these stripes get narrower.  The idea that the pseudogap is 
associated with stripe minigaps has been proposed previously\cite{TD}; the
present calculation provides a systematic doping dependence and a connection
with the VHS.

A complete understanding of the stripe phase, particularly in BSCCO, lies in the
correct inclusion of fluctuation effects.  These effects can broadly be 
separated into two categories, depending on whether the fluctuation preserves
the local density distribution or not.  In the former category fall fluctuations
in the local stripe spacing, either static or slowly varying in time, and
long-wavelength bending of the stripes.  It is likely that the energy dispersion
is a fairly localized function in space, and that these fluctuations can be 
calculated as weighted averages over the present solutions.  In this case, the
dispersion would still be a superposition of the two end phases, and the main
effect of the fluctuations would be to smear out the minigaps.  Since there is
always a gap near the Fermi level, a residual pseudogap should survive. 
Moreover, since the split-off LHB is well defined, particularly at lower
doping, it should persist at a distinguishible feature after averaging.  This
would resemble the photoemission in LSCO.

The second class of fluctuations involves fluctuations which are fast enough, or
disordered on a sufficiently short-wavelength scale so that the local density
does not lie near the two potential minima.  These fluctuations act to wipe out
the stripe fluctuations on a local level, and the question is, can they describe
the experimental results in BSCCO, where, for $T>T_c$, the two valence bands 
appear to collapse
into a single reconstructed band.  This is a plausible result: as a line
of holes fluctuates back and forth in an antiferromagnetic background, the 
background will have to adjust to some time-averaged hole density.  The 
theoretical problem is how to properly include this averaging: it is a question
of how the system responds locally on different time scales.

\subsection{Four Types of Gap}

It should be noted that the present model has four distinct types of gap in
the electronic spectra.  First, there is the Mott-Hubbard gap in the magnetic
stripes, e.g., Fig.~\ref{fig:21}a.  Then there is the Van Hove gap on the
charged stripes, Fig.~\ref{fig:18}.  It is postulated that the opening of these
two gaps provides the stabilization energy of the stripe phase, as in 
Fig.~\ref{fig:16}.  However, in 
addition there is a third type of gap, arising from the new periodicity
associated with the stripe phase.  This leads to minigaps in the dispersion
across the stripes, clearly visible in Fig.~\ref{fig:21}b-d.  Finally, there is
the superconducting gap.

Many earlier discussions of the pseudogap relied on the photoemission studies of
BSCCO in the {\it normal} state.  The recent systematic studies in 
BSCCO\cite{pdh2,pdh1} and LSCO\cite{Ino} suggest that the low-temperature data
are more representative of the stripe phase.  In this case, Fig.~\ref{fig:31} 
shows that there are two distinct features, and the Mott-Hubbard gap is 
associated with the `hump' feature in BSCCO; the `peak' feature is then 
plausibly associated with the charged stripes.  The question then is which of
the remaining three theoretical gap features accounts for the `peak'.

The doping dependence of the peak feature is not consistent with a conventional
superconducting gap: the transition temperature $T_c$ decreases while the peak 
energy $\Delta_p$
increases with increasing underdoping.  It has been suggested\cite{Zas} that
$\Delta_p$ is superconductivity-related, but associated with preformed pairs.
However, in the case of a multicomponent order parameter, it has been shown
that the total gap at $(\pi ,0)$ is a vector sum of the individual components.
In BSCCO, a careful analysis of the data\cite{MK2} (involving the detailed 
doping dependence of the `dip' feature) suggests that the gap is best 
interpreted as a superposition of a superconducting component and a 
non-superconducting component, with the latter dominating in the strongly 
underdoped limit.  Accepting this analysis, the question becomes, what is the 
origin of the non-superconducting component on the charged stripes?

The experimental observation that this peak increases as the hole doping 
decreases cannot be easily explained in terms of conventional Van Hove gaps, 
which are largest at $x_0$, Figs.~\ref{fig:16} and ~\ref{fig:18a}.  A model 
involving Van Hove pinning, introduced\cite{Pstr} 
to explain the pseudogap crossover in BSCCO, relied on the normal-state data,
which appeared to show a smooth evolution into the magnetic stripe at half 
filling.  The model cannot explain the evolution of a distinct charge-stripe
gap.  

\begin{figure}
\leavevmode
   \epsfxsize=0.44\textwidth\epsfbox{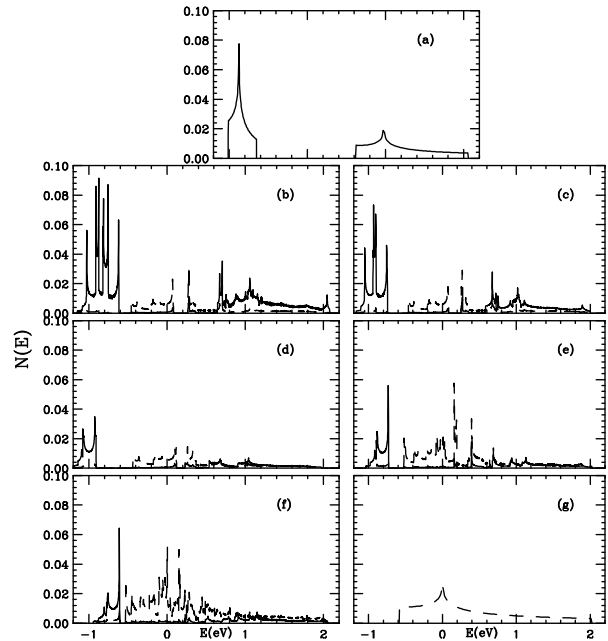}
\vskip0.5cm 
\caption{Density of states for dopings: $x$ = 0 (a), 0.0625 (b), 0.0833 (c), 
0.125 (d), 0.167 (e), 0.1875 (f), and 0.25 (g) assuming dielectric constant 
$\epsilon =15$.  (The $x=0$ data are shifted up by 0.16eV.)  Solid (dashed) 
line = partial density of states for magnetic (charged) stripes.}  
\label{fig:19}
\end{figure}

On the other hand, the minigaps seen in Fig.~\ref{fig:21} do on average grow
larger as the doping is reduced, saturating near the 1/8 crossover.
In the present, non-self-consistent calculation, these minigaps are not 
centered on the Fermi level.  However, it is clear that having $E_F$ fall in a
minigap would enhance the energy lowering, stabilizing these gapped phases and
improving agreement with experiment. 
It should be noted that the greatest energy lowering would correspond to 
having $E_F$ centered on the minigap closest to the VHS, since that is where 
the dos is highest, Fig.~\ref{fig:19}.  This would then be a form of {\it 
stripe-induced Van Hove splitting}.

\begin{figure}
\leavevmode
   \epsfxsize=0.44\textwidth\epsfbox{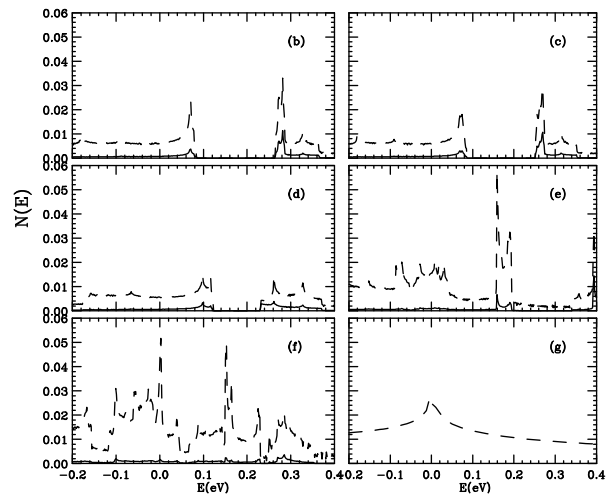}
\vskip0.5cm 
\caption{Blowup of density of states near $E_F$, for same dopings as in Figure
\protect\ref{fig:19}.  (There is no frame (a), since the dos vanishes in this
energy regime.)}
\label{fig:19b}
\end{figure}

Note that the largest minigap is associated with this VHS splitting, even in
the lowest doping situation, (6,2).  Figure~\ref{fig:19b} shows a blowup of the
dos near the Fermi level.  It can be seen that there is a gap (dos = 0) for $x
\le 0.125$), centered at an energy $\sim 0.18eV$ above $E_f$.  For larger $x$
a pseudogap (dos $>$ 0) persists, shifting toward $E_F$ with increasing doping.
Since the problem is two dimensional and
the stripes are fluctuating, this gap will be spread out into a pseudogap.  
Nevertheless, the low dos means that localization effects should be present.
Since the (pseudo)gap remains large at the 1/8 crossover, there should be a 
delocalization transition at a somewhat higher doping.  In the present 
calculation, this feature falls above the Fermi level; however, this could 
change in a fully self consistent calculation.

Finally, one should note the duality between the charge stripe minigap, which 
grows as hole doping is reduced, Fig.~\ref{fig:31}, and the magnetic stripe 
spin gap, which increases with increasing hole doping, Fig.~\ref{fig:7}. 

\subsection{Fermi Surface and Remnant Fermi Surface }

\begin{figure}
\leavevmode
   \epsfxsize=0.44\textwidth\epsfbox{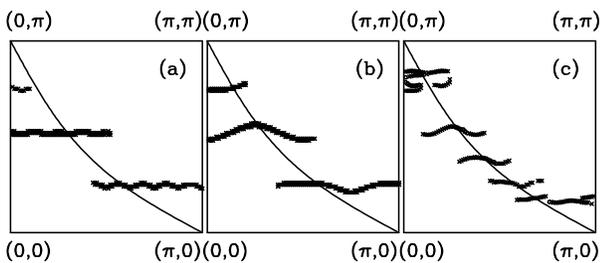}
\vskip0.5cm 
\caption{Fermi Surfaces for dopings: $x$ = 0.0625 (a), 0.125 (b), and 0.1875 
(c), and $\epsilon =15$.  Solid line in each is the Fermi surface for x=0.25.}
\label{fig:22}
\end{figure}

Figure~\ref{fig:22} shows the Fermi surfaces corresponding to the same dopings
as in Fig.~\ref{fig:21}.  As expected from the `projected' dispersions in that
Figure, in all cases the wave functions are $>90\%$ 
associated with the charged stripes.  Note the important role of the structure
factors: while there is a well-defined superlattice in each case, and hence
each Fermi surface segment is periodically repeated, the weight is highly
nonuniformly distributed, being concentrated predominantly near the limiting
x=0.25 Fermi surface (solid line in the figures).

Feng, et al.\cite{Feng} recently presented photoemission evidence for the
presence of {\it two} Fermi surfaces in BSCCO, seen at different incident photon
energies.  This result is disputed\cite{Mes}, but it is intriguing that the
`new' Fermi surface has large flat sections, very similar to the 1/8 stripe
pattern, Fig.~\ref{fig:22}b. 

Ronning, et al.\cite{oxy} introduced an alternative, well-defined energy 
surface, which they refer to as a `remnant Fermi surface' (rFs).  This is the
locus of points where the integrated photoemission intensity, taken as 
proportional to $n(k)$, falls to one half its
maximal value.  While the intensity does fall to half at the Fermi level, it
can also fall to half at an energy away from the Fermi level (N.B., the rFs is
not a surface of constant energy).  Indeed, a rFs was found for the insulating
CCOC.  We have shown that in this case the rFs (the locus of points 
where the coherence factor equals one half) maps out the superlattice zone 
boundary.  (Since the model does not include fluctuations, there is only one
band below $E_F$, and hence no photoemission distribution to integrate over.)

\begin{figure}
\leavevmode
   \epsfxsize=0.44\textwidth\epsfbox{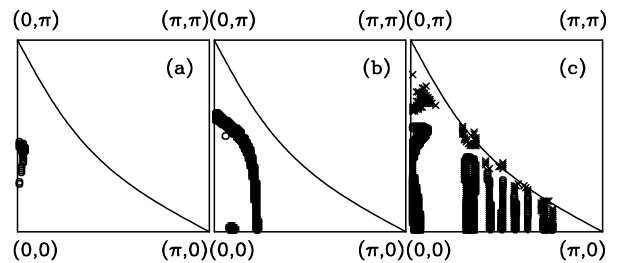}
\vskip0.5cm 
\caption{Remnant Fermi Surfaces for same dopings as in 
Fig.~\protect\ref{fig:22}: $x$ = 0.0625 (a), 0.125 (b), and 0.1875 (c).}
\label{fig:23}
\end{figure}

In the case of a stripe array, there are several complications.  First, there 
are several subbands, and one will get different results depending on whether
one calculates an rFs for each subband, or a single rFs for the whole valence
band.  The structure factor provides an additional complication, since the 
intensity is almost never the full possible value.
Nevertheless, for simplicity, Fig.~\ref{fig:23} plots the locus of points where
the net spectral function equals 1/2 -- actually, falls within a range 
0.48-0.52.  Comparison of Figures~\ref{fig:22} and \ref{fig:23} shows that the
true Fermi surface and the rFs are quite distinct features, and that the rFs 
tends to follow the superlattice Brillouin zone boundaries.

\subsection{Staging}

A test case for phase separation models of stripe formation comes from 
measurements on oxygen-doped
La$_2$CuO$_{4+\delta}$.  In these materials, the interstitial oxygens are highly
mobile, so the tendency of holes to phase separate leads to {\it macroscopic}
phase separation.  As in graphite intercalation compounds (GIC), this is 
accomplished by staging\cite{stag}: the interstitial oxygens form dense layers,
with subsequent oxygen layers separated by n layers of CuO$_2$ (n is the stage
number).  The puzzling observation\cite{BirgLSO} is that despite the macroscopic
phase separation, stripes are still found in these compounds.  I suggest the
following resolution, also based on an analogy with GIC.  The stripes were 
observed in a predominantly stage 4 compound, with 4 CuO$_2$ layers per
oxygen interstitial layer.  In GIC, a stage 4 would have a highly
inhomogeneous charge distribution, with most of the charge on the bounding
layers (in this case, the CuO$_2$ layers adjacent to the interstitial O's), and
significantly less on the interior layers, away from the O's.  Hence, in the 
cuprate I propose a similar effect: fully charged $x\simeq 0.25$ bounding
layers, and lighter charged $\sim 1/8$ interior layers.

Thus, the staging is driven by the tendency to phase separate, producing ideal
hole-doped bounding layers, but there are still stripes due to doping of the
interior layer.  This model resolves two additional problems.  First, the 
sample was a mixture of stages 2 and 4.  If the doped holes were uniformly 
distributed on all CuO$_2$ layers, then there would be {\it two independent 
stripe patterns}, since the stage 2 CuO$_2$ planes would have nearly twice as 
many holes per layer.  Within the present model, most of the charge goes into 
the bounding layer, so the layers of stage 2 would have comparable doping to the
bounding layers of stage 4; moreover, since both types of layers are nearly 
fully doped, neither would show magnetic scattering.

Finally, even though static stripe order was observed, there was no reduction in
T$_c$ -- indeed, T$_c$=42K is higher than can be obtained with Sr doping.  This
can most easily be understood if the superconductivity and the stripe order are
in different layers.

\subsection{Comparisons with Slave Boson Results}

It is instructive to compare the present results with earlier slave boson
calculations.  In the simplest version, there is no magnetic coupling, $J=0$,
and the band structure near the Mott transition is highly anomalous.  There is
a single band, but as the doping approaches half filling, $x\rightarrow 0^+$, 
the bandwidth vanishes, with both $t$ and $t^{\prime}$ renormalized to zero.  
In the three-band model, even after setting $U\rightarrow\infty$, there is
still a charge transfer energy, $\Delta$  In this case, it is also possible to
approach half filling from below, $x\rightarrow 0^-$; the same bandwidth
collapse occurs, but at a different energy, $E^-$, with $E^--E^+$ being the
(renormalized) charge transfer energy.

Is there any way to reconcile the present results with slave boson theory?  
I suggest the following possibility. 
When a hole is doped into the Mott insulator, there is phase separation, and 
locally the dispersion is restored: $t\rightarrow t_0$.  At a doping $x$, a
fraction $x/x_0$ of the electrons have hopping $\sim t$, the rest $\sim 0$.
But in the mean-field slave boson calculation the effect of the hole is
uniformly spread out over the entire lattice, leading to an effective 
$t\rightarrow xt_0$.
This is just what is found in the present stripe calculation.  As the material
is doped, the magnetic band persists with little change, while a new band
appears, characteristic of the hole-doped stripes, with full bandwidth, $t
\sim t_0$ (neglecting superlattice gaps), but with relative intensity 
proportional to $x$, Fig.~\ref{fig:21}.  If this interpretation is correct, it 
suggests that the slave boson calculation may underestimate the tendency for 
phase separation.


\subsection{Comparisons with Other Calculations}

A number of calculations\cite{Lau1,WeL,Pstr} interpret the magnetic stripe 
dispersion in terms of a flux phase rather than an antiferromagnet.  This
is unsatisfactory for two reasons: first, there is clear evidence that the
magnetic stripes have predominantly antiferromagnetic correlations; and 
secondly, the flux phase has zero gap at $(\pi /2,\pi /2)$, making it difficult
to explain the dispersion in SCOC, which has a gap $\ge 0.8eV$.  The present 
calculation is based on an antiferromagnetic model, which corrects both of 
these defects.  

The dispersion in the presence of stripes has also been calculated by 
SEK\cite{SEK} and by Seibold, et al.\cite{Sei}.  These calculations were
intended simply to show the effect of charge (or spin) modulations on the
quasiparticle dispersion and involve highly simplified models for stripes.  
SEK introduced a background stripe potential, $V_{\sigma}(\vec R)$, which fixed
the orientation and periodicity of the stripe array, and induced a modulation in
the quasiparticle charge and spin densities.  Seibold, et al. concentrate on the
charge modulation, and introduce a CDW model, again with externally imposed
periodicity.  

While these calculations demonstrated how the modulation introduces pseudogaps
into the dispersion, they are not expected to give a very good picture of
stripes arising from an underlying phase separation instability.  A CDW model
involves a {\it fixed periodicity} and a {\it variable density}, whereas phase
separation should produce an approximately fixed density and smoothly varying
periodicity.  Thus, in the present calculations, it is straightforward to
model the full doping dependence of the stripe phase; in the earlier 
calculations, the doping dependence of the periodicity and potential must be
supplied empirically, so the models can make no predictions.  Thus, neither
calculation showed any sign of the separate magnetic and charge bands.

One should note a distinction between the way the idea of CDW's is applied in 
these calculations and the usage of the present paper.  In the earlier
calculations\cite{SEK,Sei}, the charge modulation of the stripes {\it is} the
CDW.  In contrast, the present paper envisages CDW's (or related phonon
anomalies, such as the LTT phase) as existing on the charged stripes, leading
to two inequivalent sublattices -- much as antiferromagnetism leads to two
sublattices on the magnetic stripes.

In a recent paper, Pryadko, et al.\cite{PK} presented a detailed criticism of
the early Hartree-Fock calculations of stripes.  In these calculations, the 
stripes are claimed to arise from Fermi surface nesting, so some of the 
criticisms apply to a larger class of models.  Here, I would like to take issue
with some of the statements.  First, Pryadko, et al. state ``the density of 
holes along a stripe varies continuously as a function of $x$'', citing Yamada,
et al.\cite{Yam}.  While this is mainly associated with the saturation of the
incommensurability above $x$=1/8, they note that the data at lower doping also
show some curvature, suggestive of a `slight variation' of the doping.  This is 
mainly associated with the cutoff of 
incommensurability below $x\simeq 0.06$, but that data have now been superseded
by better samples: the stripes are present, but rotated by 45$^o$, and the
incommensurability falls close to the Yamada (straight) line.  As discussed 
above, both a slight doping-dependence below $x=1/8$, and a stronger saturation
above that doping {\it can} be understood in a model with a prefered charged
stripe density, in the presence of either long-range Coulomb repulsion (which
can produce a {\it real} saturation of the incommensurability) or 
commensurability pinning (which produces an {\it apparent} saturation).

Pryadko, et al. further state, ``In the LSCO family, ..., there is simply no
vestige of a quasiparticle in the region of momentum space where the nested
Fermi surface is supposed to occur.''  This statement raises several issues.
First, one must carefully distinguish conventional Fermi surface nesting from
Van Hove nesting.  For the conventional nesting model\cite{LSKL} they cite, the
nesting is associated with flat sections of Fermi surface; in LSCO these flat
sections arise near\cite{KSL} $(\pi /2,\pi /2)$, where there {\it are} 
well-defined quasiparticles.  Conversely, the fact that these quasiparticles 
persist in the striped phase shows that they are not involved in the nesting 
process.

In contrast, in Van Hove nesting the important quasiparticles are those near
$(\pi ,0)$.  In all the cuprates, these quasiparticles are found to be shifted
below the Fermi level, and indeed {\it a comparison of photoemission and 
tunneling studies reveals that the position of these quasiparticles defines the 
pseudogap} -- 
exactly as required by a Van Hove nesting theory.  Moreover, even in the SEK
calculation it was clear that these Van Hove quasiparticles, which constitute
the `flat bands' remain well defined in the calculated striped phase.  This is
further clear from, e.g., Fig.~\ref{fig:21} above.  True the quasiparticles are
greatly smeared out, but this is a direct consequence of the formation of
stripes, and cannot be taken as an argument against this mechanism of stripe 
formation.

Finally, Stojkovi\'c and Pines\cite{StoP} claimed to rule out the VHS in the
physics of the cuprates.  They employed parameters for which the VHS falls at
very high doping, $x=0.55$, and concluded ``the presence of the van Hove 
singularities near the Fermi surface plays only a marginal role".  However,
their more recent calculations\cite{SPS} employ a revised parameter set (i.e.,
$t^{\prime}=-0.25 t$), which places the VHS at $x=0.22$, close to the present 
$x_0=0.25$. 

\section{Conclusions}

This has been a long manuscript, which presents a coherent view of the stripe
phases in the cuprates.  A number of principal results of the calculations are
here summarized.  Most of the results are {\it generic}, and would be expected
in any model where the stripes result from two-phase coexistence, while a few
are specific to a Van Hove scenario.

(1) This is the first calculation of photoemission in the stripe phase involving
strong stripe correlations, i.e., preferred hole dopings with independent
characteristic dispersions, as opposed to a single dispersion with sinusoidal
(CDW/SDW) modulations.

(2)  This allows a study of the evolution of the dispersion as a function of
hole doping.
 
(3) It is found that, even at this nanoscale level, the dispersion can be
characterized as a {\it superposition of two components}, leading to a picture 
of magnetic stripe bands and hole-doped (`charged') stripe bands.  This allows a
natural interpretation of the photoemission spectra in LSCO, and suggests a
unified picture with BSCCO and SCOC.

(4)  The calculations suggest that an important role of the superconducting 
transition is to freeze out fluctuations of the stripes.  This freezeout 
manifests itself in three ways: (a) the electron-electron scattering rate drops
by several orders of magnitude below T$_c$\cite{eesup}; (b) the
photoemission dispersion splits in BSCCO into a characteristic peak-dip-hump
structure; (c) the $(\pi ,\pi )$ magnetic neutron scattering in YBCO sharpens
below T$_c$, revealing a characteristic spin gap.

(5) The doping dependence naturally leads to a picture of a series of {\it 
quantum critical points} (QCP's) or magic dopings, at which the stripe pattern 
is commensurate with the crystalline lattice.  The most prominent one is the 
famous 1/8 effect, but the metal-insulator transition in LSCO and the onset of 
superconductivity are close to two other magic numbers.

(6) The percolation crossover at 1/8 doping provides a simple model of the spin
gap in YBCO, showing that a two-leg ladder provides a good model for an
isolated magnetic stripe.  

(7) As a result of point (3), the model has a {\it natural VHS pinning} to the
Fermi level: if the VHS is at the Fermi level in the charged stripe end phase
(as it must be, if this phase is stabilized by Van Hove nesting), then the VHS
remains close to $E_F$ over the entire doping range.

(8) This provides a new explanation of the pseudogap: {\it stripe-induced
Van Hove splitting}.

(9) More speculatively, since superconductivity in YBCO is strongest well
beyond the percolation crossover (1/8 effect), {\it superconductivity seems
to be a property predominantly of the charged stripes}.

There are a number of advantages of the present model of fractionally-doped
stripes.  First, if the stripes are stabilized by CDW formation, then there is
an important continuity between stripes in the cuprates, and those in the 
nickelates and manganites.  Such continuity is lost in the SO(5) model, where
the charged stripes are stabilized by superconductivity.  Moreover, a connection
with CDW's would naturally explain the experimental observation that the stripe 
phases are dominated by charge order rather than spin order, a result difficult
to understand in a pure Hubbard or tJ model.

{\bf Acknowledgment:}  These computations were carried out using the facilities
of the Advanced Scientific Computation Center at Northesatern University.  
Their support is gratefully acknowledged.  I thank C. Kusko, S. Sridhar, A.
Bansil, B. Barbellini, and M. Lindros for stimulating conversations.
Publication 776 of the Barnett Institute.

\appendix

\section{White-Scalapino Stripes}

It remains an open question whether phase separation is a generic feature of
the Hubbard and tJ models, or arises only in a restricted parameter domain.  
In the tJ model, the recent density-matrix renormalization group (DMRG) 
calculations of White and Scalapino (WS)\cite{WhiSc,WhiSc2} find clear evidence 
for stripes, but even these results are controversial: some groups find
that stripes are metastable within the tJ model\cite{Put,HelMan} (see the 
discussion in Refs.\onlinecite{WS3,HM2}), others that the model should display 
macroscopic phase separation\cite{pry}.  Moreover, inclusion of realistic values
of $t^{\prime}$ into the model further reduces the stripe 
stability\cite{Tohy,WS4}.  Nevertheless, they represent at worst a low-lying 
excited state, and hence may become the true ground state in the presence of 
additional interactions (e.g., electron-phonon coupling).  

It is interesting to note that many of the DMRG results can be simply understood
in terms of a phase separation model. 
This hypothesis can explain four quantitative results of WS: (1) At 1/8 doping, 
WS\cite{WhiSc} find hole-doped stripes which are 2 Cu atoms wide, with doping 
$\sim 0.25$ holes per Cu, which they interpret as a bond-centered domain wall 
with 0.5 holes per unit length.  Increasing the doping, (2) they find a 
remarkable transition to stripes with 1 hole per unit length\cite{WhiSc2}, 
(3) which coexist with the 0.5 hole stripes for 0.125$\le x<0.17$, and finally, 
(4) the stripe phase disappears at x=0.3. 

All of the above features can be simply understood by assuming that the 
hole-doped stripes have a density $x^*\sim 0.25$ hole/Cu, and that there 
is a preference for both magnetic and charged stripes to have even width.  
Hence, the narrowest possible magnetic stripes are two cells wide.

Given these assumptions, the WS data can be explained as follows.  In the low
doping regime, hole doped stripes form with the minimal width of two cells, and
0.25 holes per Cu, for a net of 0.5 holes per unit length.  This continues up
to 1/8 doping, at which point both magnetic and hole-doped stripes are two
cells wide.  Doping holes beyond 1/8 doping can be accomplished by increasing 
the width of the hole-doped stripes.  A stripe 4 Cu wide has 4$\times$0.25 = 1 
hole per unit length.  The sample will be a mix of 2-Cu and 4-Cu stripes until
$x=2x^*/3\sim 0.167$, and when $x=x^*\sim$0.25, the whole sample is filled 
with the hole-doped phase, and the stripe phase terminates.
This last number is lower than the simulation, $x=0.3$, and may be evidence
that $x^*$ is weakly doping dependent. 

\begin{figure}
\leavevmode
   \epsfxsize=0.33\textwidth\epsfbox{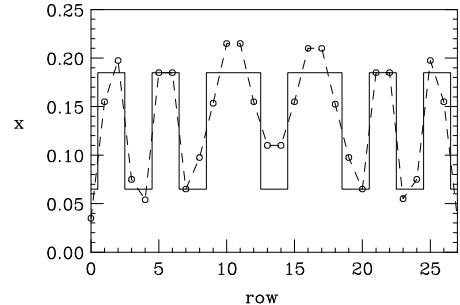}
\vskip0.5cm 
\caption{Rung-averaged hole density (open circles)\protect\cite{WhiSc2}, 
interpreted as constant hole density domains of variable width.  
Dashed lines = guides to the eye; solid lines = phase separation model.}
\label{fig:1b}
\end{figure}

The WS simulations actually display this increase in stripe width 
above 1/8 doping, Fig. 4a of Ref.~\cite{WhiSc2}.  This is more clearly seen in 
plots of the rung-averaged hole density, Fig.~\ref{fig:1b}.  The data (open 
circles) fall very close to the form expected for a phase separation model 
(solid lines), with the same average densities as at 1/8 filling, but now the 
magnetic stripes retain their minimum width, while the hole-doped stripes get 
wider. 

\begin{figure}
\leavevmode
   \epsfxsize=0.33\textwidth\epsfbox{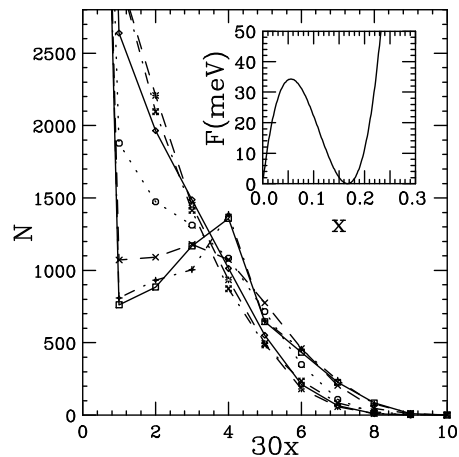}
\vskip0.5cm 
\caption{Striped phase melting transition at x=0.04, $\epsilon =23$: hole 
distribution function at temperatures (reading from bottom to top, at fixed
hole occupancy $30x=2$) $k_BT$ 
= 10, 0.1, 30, 60, 100, 400, 200 meV.  Inset: Assumed free energy vs. doping.}
\label{fig:C1}
\end{figure}

It should be stressed that the mechanism responsible for stabilizing the WS 
stripes is unknown, but clearly does not involve the VHS, being weakened by a
finite $t^{\prime}$.  Even should the WS stripes prove to be the ground state,
it may still be that they are too weakly bound to explain experimental 
observations.  The debate on their stability suggests that non-striped phases 
lie very close in energy -- for instance, Hellberg and Manousakis\cite{HM2} 
find energy differences of the order of 1meV.  But it is found that the stripe 
phase separation collapses when the temperature becomes comparable to the energy
barrier between the phases\cite{MV}, Fig.~\ref{fig:C1}.  
It should be noted that
Figure 2 of Ref.~\onlinecite{MV} was printed incorrectly in the original 
publication.  The correct figure is included here, as Fig.~\ref{fig:C1}. 
The calculations are described in the original publication.  Note that, at low
temperatures, the particle distribution is two peaked, corresponding to the
two kinds of stripes, while at high temperatures the distribution collapses into
a monotonic limit.  The crossover temperature scales with the free energy 
barrier between the two phases.

\section{More on the SDW Dispersion}

\begin{figure}
\leavevmode
   \epsfxsize=0.33\textwidth\epsfbox{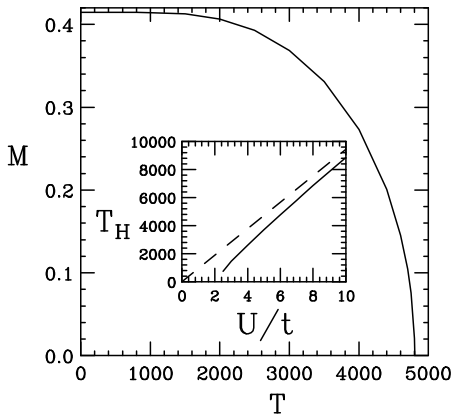}
\vskip0.5cm 
\caption{Temperature dependence of magnetization in mean-field model.  Inset:
$U$-dependence of $T_H$.  Dashed line: $T=U/4$.}
\label{fig:B1}
\end{figure}

As will be shown below, the SDW model gives a surprisingly good description of
the photoemission dispersion, not only in the insulating limit, but also in 
doped BSCCO.  However, if $M$ is interpreted as the long-range N\'eel order
parameter, there are problems with the doping dependence, Fig. \ref{fig:1}, and
the temperature dependence, Fig. \ref{fig:B1}.  On the other hand, since this is
a mean-field calculation, it is more plausible to interpret $M$ as a measure of
short-range antiferromagnetic order, and the transition as describing the
splitting of the upper and lower Hubbard bands.  To stress this point, the
transition is labelled as $T_H$.  Figure~\ref{fig:B1} shows the temperature 
dependence of the magnetization, while the inset shows the dependence of $T_H$ 
on $U/t$.  These results are consistent with the three-dimensional calculations 
of Dichtel, et al.\cite{DJK}.

Interpreting the doping dependence of $M$ as a measure of the renormalization 
of the {\it splitting} into upper and lower Hubbard bands, the mean-field 
calculations are in good agreement with exact
diagonalization calculations\cite{EsO} and experiments on the 
cuprates\cite{exp}.  
 
\begin{figure}
\leavevmode
   \epsfxsize=0.33\textwidth\epsfbox{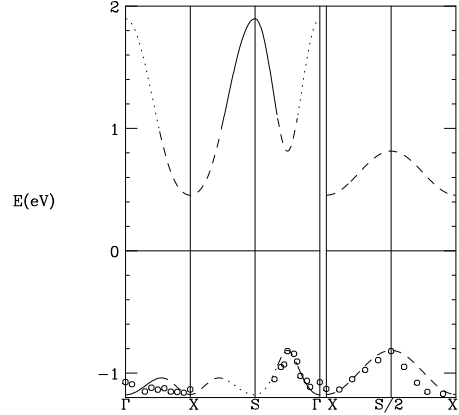}
\vskip0.5cm 
\caption{Dispersion of the antiferromagnetic insulator in mean-field model.  
Open circles = data of Ref.~\protect\cite{Well}.  Solid lines: coherence factor
$>0.8$; dashed lines: $0.8>$ coherence factor $>0.2$: dotted lines: coherence
factor $<0.2$.}
\label{fig:B2}
\end{figure}

\begin{figure}
\leavevmode
   \epsfxsize=0.33\textwidth\epsfbox{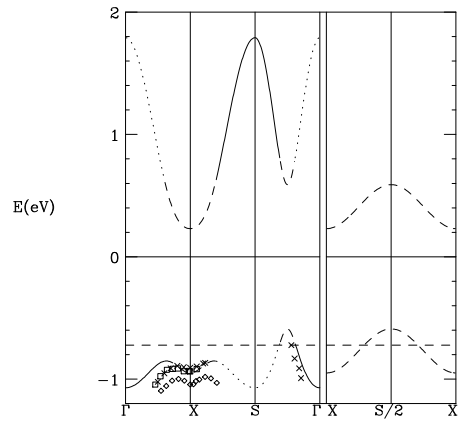}
\vskip0.5cm 
\caption{Dispersion of the doped antiferromagnet, $x=0.169$, in mean-field 
model.  $\times$'s = data of Ref.~\protect\cite{Gp0} (underdoped, $T_c=67K$); 
diamonds (underdoped, $T_c=52K$) and squares (overdoped, $T_c=72K$) = data of 
Ref.~\protect\cite{pdh2}; solid lines: coherence factor $>0.8$; dashed lines: 
$0.8>$ coherence factor $>0.2$: dotted lines: coherence factor $<0.2$.  
Horizontal dashed line = Fermi level.}
\label{fig:B3}
\end{figure}

\begin{figure}
\leavevmode
   \epsfxsize=0.33\textwidth\epsfbox{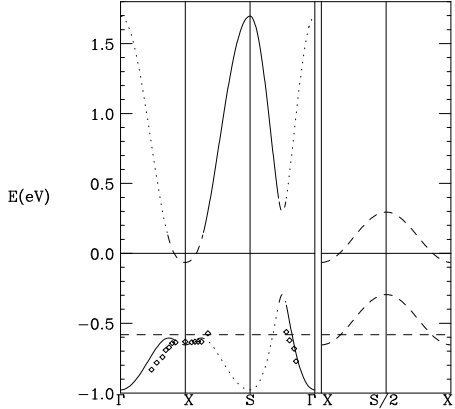}
\vskip0.5cm 
\caption{Dispersion of the doped antiferromagnet, $x=0.356$, in mean-field 
model.  Diamonds = data of Ref.~\protect\cite{Gp0} (overdoped, $T_c=85K$); 
solid lines: coherence 
factor $>0.8$; dashed lines: $0.8>$ coherence factor $>0.2$: dotted lines: 
coherence factor $<0.2$.  Horizontal dashed line = Fermi level.}
\label{fig:B4}
\end{figure}

\begin{figure}
\leavevmode
   \epsfxsize=0.33\textwidth\epsfbox{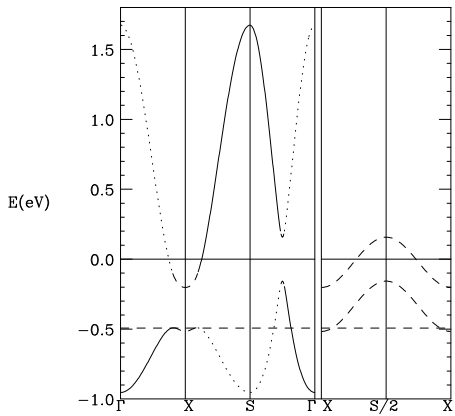}
\vskip0.5cm 
\caption{Dispersion of the doped antiferromagnet, $x=0.392$, in mean-field 
model.  Solid lines: coherence 
factor $>0.8$; dashed lines: $0.8>$ coherence factor $>0.2$: dotted lines: 
coherence factor $<0.2$.  Horizontal dashed line = Fermi level.}
\label{fig:B5}
\end{figure}

Figures \ref{fig:B2}-\ref{fig:B5} show how the dispersion changes with doping.  
For all curves, the same band parameters, given below Eq. \ref{eq:A2}, are
used, with $M(x)$ found self consistently.  Surprisingly, the results are 
qualitatively consistent with the pseudogap in the underdoped cuprates 
($\times$'s, diamonds, and squares)\cite{Gp0,pdh2}, even though no striped
phases are assumed.  However, the predicted hole doping is anomalously large.

In the undoped material, the model has a direct (indirect) 
Mott-Hubbard gap of 1.63eV (1.27eV), and reproduces the dispersion found in the 
oxyclorides\cite{Well,oxy}, Fig.~\ref{fig:B2}.  Here $X=(\pi ,0)$, $S=(\pi ,\pi 
)$, and $\bar S=S/2$.  The spectral weight is proportional to the coherence 
factor
\begin{equation}
\zeta_{\pm}={1\over 2}\bigl(1\pm{\epsilon_-\over W}\bigr),
\label{eq:A2b}
\end{equation}
with the subscript $+$ ($-$) referring to the upper (lower) Hubbard band.  From
Fig.~\ref{fig:B2} it can be seen that the coherence factor is $\le 0.5$ all 
along the $X\rightarrow S$ branch, where no quasiparticle peaks were reported.
The weight along the line $X\rightarrow\bar S$ is uniformly 1/2,
characteristic of the antiferromagnetic zone boundary\cite{BSW,KM}.

In studies of the Fermi surfaces of the hole-doped cuprates, there is an
important question as to whether the Fermi surfaces are large, centered on
$\Gamma$, or small hole pockets, centered at $\bar S=(\pi /2,\pi /2)$, and 
equivalent points.  In dispersion studies, the hole pockets are associated with
retrograde dispersion branches of low intensity (ghosts), particularly along
$X\rightarrow S$ and $S\rightarrow\bar S$.  Note that this negative curvature 
along the branch $X\rightarrow S$ is clearly seen for underdoped samples with 
$T_c\le 75K$\cite{Gp0,pdh2}, Fig.~\ref{fig:B3}.  
[Parenthetically, the striking difference in doping for the two data sets in 
this figure arises because Marshall, et al.\cite{Gp0} are plotting
the normal state pseudogap, while Campuzano, et al.\cite{pdh2} plot the hump 
feature in the superconducting state.]

For these parameters, the transition to the paramagnetic phase is first order; 
for other parameter choices, a second-order transition is found.  The first
order transition is a topological transition, arising when the Fermi level
crosses the band dispersion near $X$ (Fig.~\ref{fig:B5}).  This is rather 
striking, since topological transitions are typically rather weak -- of order
2.5.  A similar result was found in the Hubbard model ($t^{\prime}=0$)
by Guinea, et al.\cite{GLLV}.

Thus, the dispersion of the normal state pseudogap, as seen in photoemission, 
can be well described by a uniform doping of the magnetic phase, in the absence 
of stripes.  This has been noted by a number of groups, including Schmalian, et 
al.\cite{SPS}, and Misra, et al.\cite{MiG}.  
However, this cannot be the complete interpretation.  Tunneling 
studies clearly demonstrate that the pseudogap consists of two components, 
split approximately symmetrically about the Fermi level.  The uniform phase 
dispersion describes the feature {\it below} the Fermi level, but the model 
contains no corresponding feature above the Fermi level.

\end{document}